\PassOptionsToPackage{table,dvipsnames}{xcolor}
\documentclass[11pt]{cuhksz}
\usepackage{url}
\usepackage{tabularx}
\usepackage{amsmath,amssymb}
\usepackage{pifont}
\usepackage{array}
\usepackage{wrapfig}
\usepackage{needspace}
\usepackage{float}
\usepackage[utf8]{inputenc}
\usepackage{CJKutf8}
\pretocmd{\section}{\Needspace{6\baselineskip}}{}{}
\pretocmd{\subsection}{\Needspace{5\baselineskip}}{}{}

\newcommand{\yes}{\ding{51}}
\newcommand{\no}{\textcolor{gray}{\ding{55}}}
\newtcolorbox{promptbox}[1][]{
  breakable,
  colback=gray!5,
  colframe=gray!45,
  boxrule=0.5pt,
  arc=1.5pt,
  left=5pt, right=5pt, top=4pt, bottom=4pt,
  fonttitle=\bfseries,
  fontupper=\small\ttfamily,
  #1
}

\leftlogo{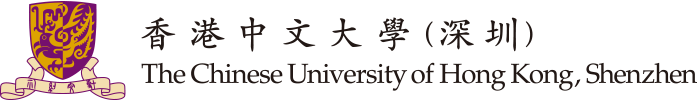}
\makeatletter
\gdef\cuhksz@rightlogos{%
}
\makeatother

\title{\texorpdfstring{\raisebox{-0.2\height}{\includegraphics[height=1.4em]{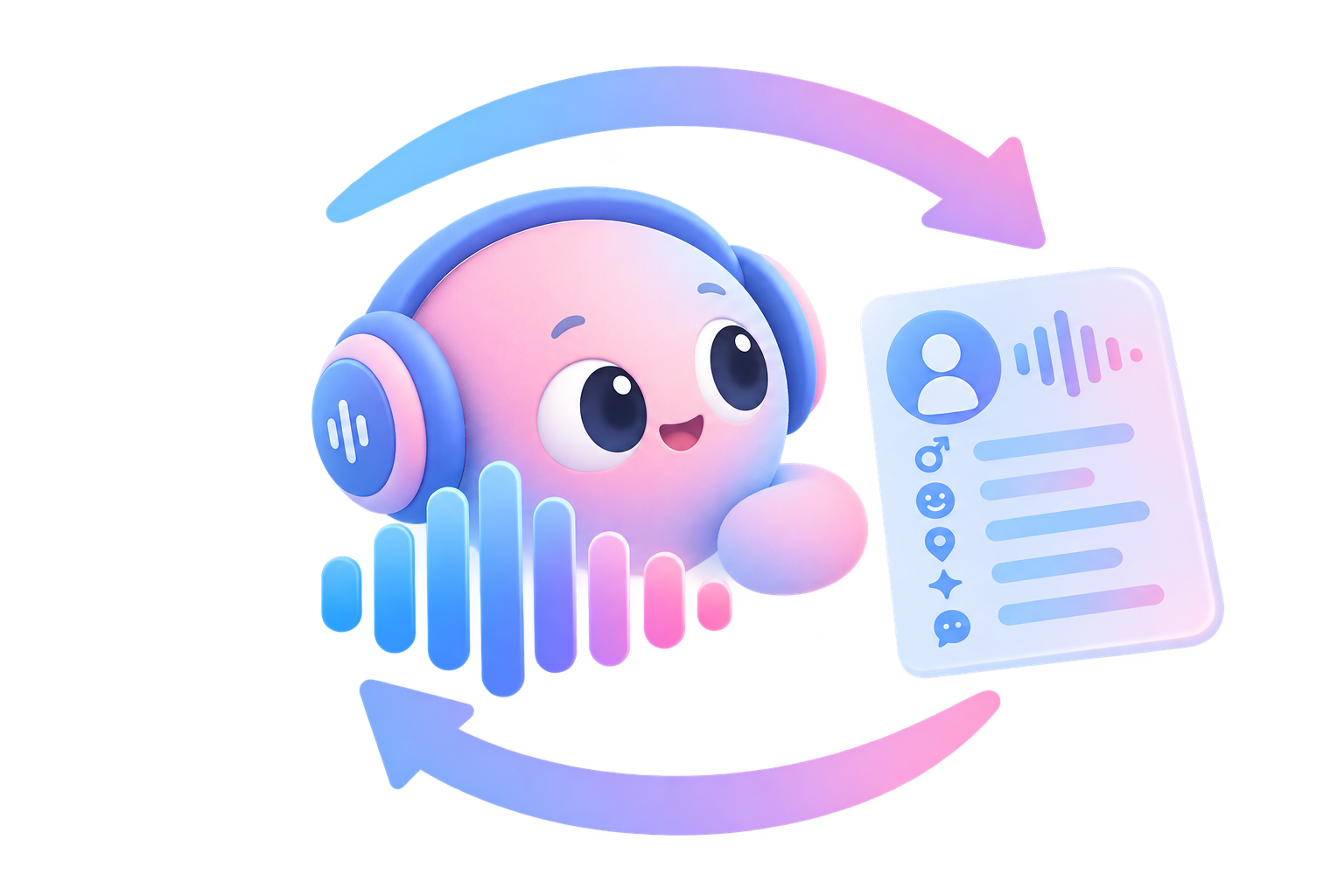}}\hspace{0.3em}}{}CycleSpeech: Reciprocal Alignment for Instruction-\\
Controlled Speech Synthesis and Paralinguistic Understanding}
\author[1]{Huan Liao}
\author[2]{Haonan Han}
\author[1]{Xingwen Han}
\author[1]{Dekun Chen}
\author[1]{Yuancheng Wang}
\author[1,3,\dagger]{Zhizheng Wu}

\contribution[\dagger]{Corresponding author.}
\affiliation[1]{The Chinese University of Hong Kong, Shenzhen}
\affiliation[2]{Tsinghua University}
\affiliation[3]{Amphion Technology Co., Ltd.}

\hypersetup{
  pdftitle={CycleSpeech: Reciprocal Alignment for Instruction-Controlled Speech Synthesis and Paralinguistic Understanding},
  pdfsubject={Instruction-controlled speech synthesis and paralinguistic understanding},
  pdfauthor={}
}

\abstract{
Instruction-controlled speech synthesis and paralinguistic understanding are often trained independently, leaving reciprocal feedback between the two tasks underexplored. We introduce \textbf{CycleSpeech}, a framework that connects generation and understanding through a shared, structured \emph{voice profile} that serves as a common target for supervision and reciprocal feedback. The forward cycle assesses whether synthesized speech expresses the intended attributes by comparing recovered and target profiles. The backward cycle evaluates whether profiles inferred from real speech can guide reconstruction of the source speaking style. To support both directions, we construct a bilingual dataset of 20,046 examples pairing instructions, target speech, speaker references, and structured profiles. Building on joint supervised fine-tuning, \textbf{CycleGRPO} alternates policy updates using reciprocal rewards grounded in profile consistency and speaking-style reconstruction. Fixed target profiles anchor feedback from the evolving counterpart. This procedure requires neither human preference annotations nor an additional preference-trained reward model. Evaluations on Chinese and English benchmarks show improved instruction adherence and profile recovery while maintaining competitive synthesis quality. Compared with Step-Audio-2-mini, CycleSpeech improves instruction-match accuracy by 4.50 and 10.06 percentage points in Chinese and English, respectively. Controlled ablations further support the contribution of cycle feedback to generation control. These results support structured voice profiles as an interface for reciprocal training between speech generation and paralinguistic understanding. An online demo is available at \url{https://cyclespeech.github.io}.
}

\begin{document}
\maketitle
\makeabstract

\section{Introduction}

Instruction-controlled text-to-speech (TTS) aims to realize user-specified
vocal attributes while preserving naturalness and intelligibility.
Text-based control has expanded from predefined style descriptions
\cite{guo2023prompttts,yang2024instructtts,ji2024textrolspeech} to open-ended
instructions \cite{ren2026ovinstructtts,zhou2024voxinstruct,yang2025emovoice},
sometimes combined with speaker references
\cite{chen2026flexivoice,ji2024controlspeech}.
Yet natural-sounding speech can still miss the requested emotion, vocal
texture, or delivery. The challenge is therefore not only to interpret an
instruction, but also to verify that its attributes are audible in the output.

Recent approaches make the intended vocal delivery explicit before synthesis.
OV-InstructTTS \cite{ren2026ovinstructtts} infers emotion labels, acoustic
descriptions, and paralinguistic tags before synthesis. BatonVoice
\cite{wang2026batonvoice} uses an external language model to plan vocal
features for a separate TTS model. Such planning specifies what the
synthesizer should express, but does not by itself require the resulting
speech to recover the intended attributes. Conversely, a description
inferred from speech may match an annotation without preserving enough
information to reproduce the source's speaking style.

We therefore ask: \emph{Can speech generation and paralinguistic
understanding provide useful training feedback to each other?}
We introduce \textbf{CycleSpeech}, which uses a structured \emph{voice
profile} as a common target for understanding supervision and generation
verification. The profile describes speaker traits, prosody, affect, and
context in a canonical schema, making recovered attributes comparable
with their targets. Generation still takes natural-language instructions
as input; the target profile provides training supervision rather than
an additional inference-time requirement.

\begin{wrapfigure}[18]{r}{0.49\textwidth}
    \centering
    \includegraphics[width=\linewidth]{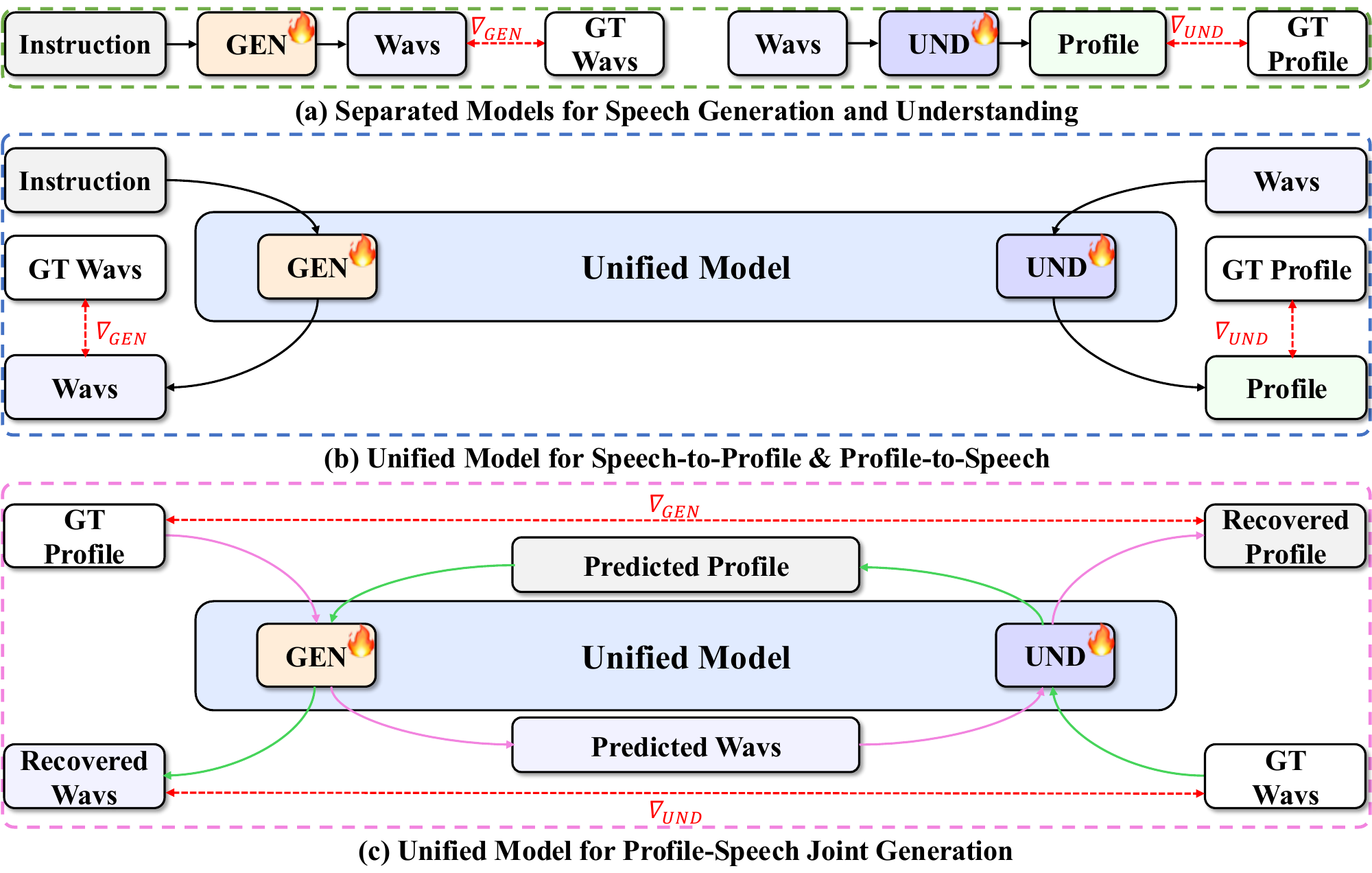}
    \caption{Generation--understanding paradigms. (a) Separate models optimize generation and understanding independently. (b) A shared backbone supports both tasks but still receives one-way supervision. (c) CycleSpeech closes reciprocal profile--speech cycles, allowing generation and understanding to verify each other.}
    \label{fig:cyclespeech_overview}
\end{wrapfigure}
The shared profile connects the tasks through two complementary training
checks. In the forward loop, understanding recovers a profile from
generated speech and compares it with the target, supplying feedback on
realized control. In the backward loop, generation reconstructs speech
from an inferred profile, testing whether the description preserves the
source's speaking style. Together, these loops evaluate profiles through
both agreement with annotations and their consequences for synthesis.
Figure~\ref{fig:cyclespeech_overview} contrasts this reciprocal design
with separate and parameter-shared systems.

Reciprocal training requires a shared representation that links generation
targets to attributes inferred from speech. Existing datasets pair speech with
style attributes or natural-language captions
\cite{ji2024textrolspeech,jin2024speechcraft,diwan2025paraspeechcaps,wang2025capspeech}, but lack a canonical profile aligned with both instructions and speech for bidirectional supervision.
Our dataset comprises 20k bilingual examples, each aligning a target utterance
with its transcript, speaker reference, instruction, and voice
profile. This profile enables reference-based alignment: generated speech is
mapped back to a profile, and categorical agreement with the target yields a
verifiable optimization signal. It also supervises speech understanding and
guides speaking-style reconstruction.

We first establish a compatible initialization through joint supervised
fine-tuning (Joint SFT), adapting understanding to generated speech after
a generation warm-up. We introduce \textbf{CycleGRPO}, which alternately optimizes generation
and understanding through reciprocal feedback. The understanding branch
provides unified profile-verification feedback without separate
attribute-specific reward classifiers, while training requires neither
human preference annotations nor a separately trained reward model.
Our contributions are:
\begin{itemize}
\item We introduce CycleSpeech, connecting instruction-controlled
speech synthesis and paralinguistic understanding through a shared,
recoverable voice profile. The forward cycle verifies whether generated speech realizes the
intended attributes, while the backward cycle tests whether inferred
profiles support reconstruction of the source speaking style.
\item We develop CycleGRPO to optimize both cycles through alternating
policy updates, with evolving counterparts providing feedback and
fixed profile targets anchoring learning. It preserves natural-language
generation inputs and requires neither human preference annotations
nor a separately trained preference reward model.
\item We construct a bilingual instruction--speech--profile
dataset that jointly supports understanding supervision and generation
verification. Bilingual evaluations and controlled ablations demonstrate the importance
of reciprocal cycle feedback, with improved instruction following and
categorical profile recovery over no-cycle and fixed-feedback baselines.
\end{itemize}
\section{Related Works}
\subsection{Instruction-Controlled Speech Generation}

PromptTTS~\cite{guo2023prompttts}, InstructTTS~\cite{yang2024instructtts},
Parler-TTS~\cite{lyth2024parlertts}, and Audiobox~\cite{vyas2023audiobox}
condition synthesis on textual descriptions.
ControlSpeech~\cite{ji2024controlspeech} and
VoxInstruct~\cite{zhou2024voxinstruct} combine instruction control with
zero-shot voice cloning. Recent methods add open-vocabulary reasoning,
vocal-feature planning, or joint style--timbre
control~\cite{ren2026ovinstructtts,wang2026batonvoice,zhang2025vevo}.
Post-training further refines synthesis through task-specific rewards:
CosyVoice~3~\cite{du2025cosyvoice3} uses ASR rewards, while Fish Audio~S2~\cite{liao2026fishaudios2}
uses ASR, quality, and speaker feedback.
FlexiVoice~\cite{chen2026flexivoice} progresses from emotion-focused DPO and
GRPO to complex instruction following guided by audio-language-model rewards.
These methods optimize synthesis, whereas CycleSpeech jointly aligns generation
and paralinguistic understanding through profile verification and speaking-style
reconstruction. Attribute- and caption-based
datasets~\cite{ji2024textrolspeech,ji2024controlspeech,jin2024speechcraft,diwan2025paraspeechcaps,wang2025capspeech}
provide descriptive supervision. Our corpus further supplies canonical profiles
as shared targets for understanding supervision and generation verification.

\subsection{Verifiable Shared Targets for Speech Generation and Understanding}

Connecting generation and understanding requires a shared target that supports both generation and verification. Structured descriptions provide such an interface by mapping instructions to explicit paralinguistic attributes that can be recovered from speech.
SpeakerCard-1M~\cite{peng2026speakercard} organizes speaker and utterance traits
into a typed schema, while UniStyle~\cite{zhu2024unistyle} supports both
speaking-style captioning and synthesis through shared text descriptions.
However, its two directions remain independently supervised. CycleSpeech instead
uses a structured voice profile as an auditable interface for understanding and generation, making reciprocal
verification well defined.

\subsection{Paralinguistic Speech Understanding and Evaluation}

External evaluators and understanding models can judge or describe speaking
style. InstructTTSEval~\cite{huang2025instructttseval} and audio-aware LLM
judges~\cite{chiang2025audioaware} assess instruction--speech consistency,
whereas StyleCap~\cite{yamauchi2024stylecap} and
ParaSpeechCLAP~\cite{diwan2026paraspeechclap} caption or align broader style
semantics.
Speech-chain methods jointly train ASR and TTS by reconstructing
speech, text, or intermediate representations~\cite{tjandra2017speechchain,hori2019cycle}.
Cycle consistency has also been used to preserve reference styles
during TTS style transfer~\cite{whitehill2020cycle,xue2021cycle}. Other work introduces
speaker-consistency constraints and staged optimization to prevent
TTS from overfitting to the ASR objective~\cite{makishima2022speaker}.
CycleSpeech extends this principle to an explicit, recoverable voice profile
linking instruction-controlled synthesis and paralinguistic understanding.
Attribute verification and speaking-style reconstruction supply reciprocal
feedback for alternating policy updates in CycleGRPO, while generation
retains its natural-language instruction interface.

\section{Method}
\begin{figure*}[t]
    \centering
    \includegraphics[width=\linewidth]{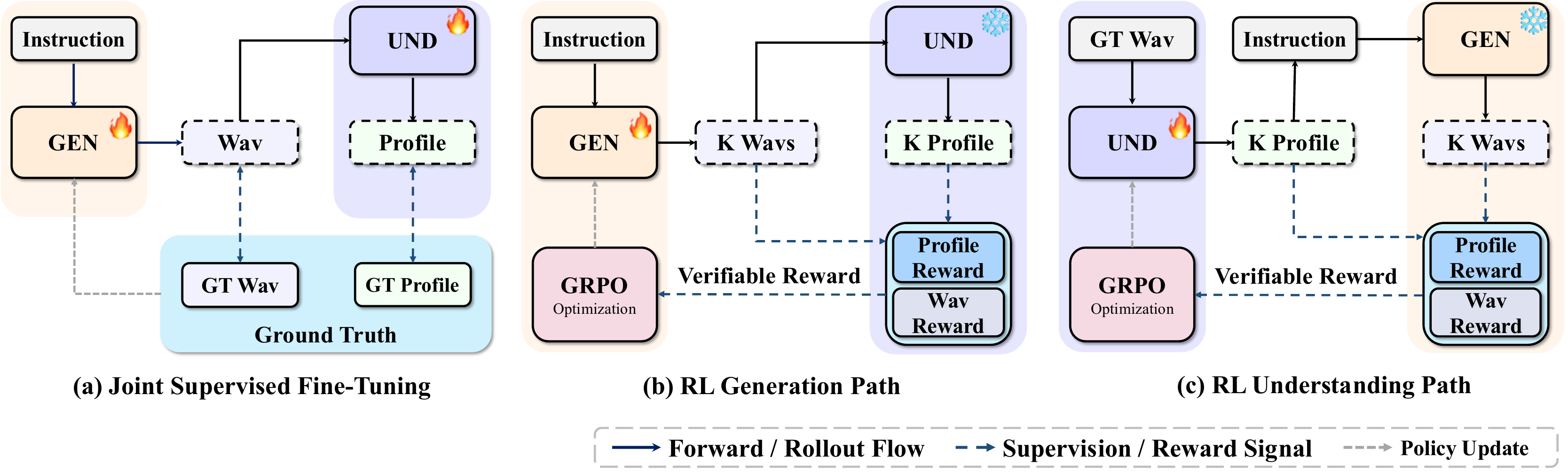}
    \caption{CycleSpeech training framework. (a) Joint SFT initializes generation (GEN) and understanding (UND) with paired speech--profile supervision. (b,c) CycleGRPO alternately optimizes generation and understanding through reciprocal verifiable rewards.}
    \label{fig:cyclespeech_pipeline}
\end{figure*}

\subsection{Overview}
CycleSpeech couples instruction-controlled generation and paralinguistic
understanding through a structured voice profile (Figure~\ref{fig:cyclespeech_pipeline}).
GEN takes an instruction, transcript, and speaker reference waveform as inputs.
The gold profile provides supervision and verification targets, rather
than an additional conditioning input to GEN.
A frozen Step-Audio-2-mini~\cite{wu2025stepaudio2technicalreport} backbone $\mathcal{M}_{\phi}$ is shared by two LoRA
adapters: GEN, parameterized by $\theta_G$, and UND, parameterized by
$\theta_U$. Training comprises Joint SFT for compatible bidirectional
initialization and CycleGRPO for reciprocal alignment. The waveform and profile
interfaces are discrete, so only the active adapter receives gradients in each
optimization direction.

\subsection{Training Data Construction}
\label{sec:data_construction}
Each training instance contains an instruction $I$, transcript $T$, speaker
reference $R$, target speech $x^{*}$ with discrete tokens $A^{*}$, and
canonical voice profile $P^{*}$. The same profile serves as the supervision target for UND and the verification target for GEN, enabling bidirectional training through these aligned annotations. Following
InstructTTSEval~\cite{huang2025instructttseval}, instructions span three
categories: Acoustic-Parameter Specification (APS) for detailed acoustic
control, Descriptive-Style Directive (DSD) for high-level style descriptions,
and Role-Play (RP) for contextual scenarios and character traits.

\paragraph{Profile and prompt construction.}
We curate human speech from WenetSpeech, a multi-domain Mandarin corpus spanning diverse speaking styles, and an internal collection of English film and television recordings. After filtering music, non-speech,
and multi-speaker segments, Gemini-3-flash~\cite{googledeepmind2025gemini3flash}
produces a qualitative report covering speaker characteristics, vocal texture,
prosody, emotion, and inferred speaking context. DeepSeek-v3~\cite{liu2024deepseek}
for Chinese and Gemini-3-flash for English then normalize these reports into
a shared JSON schema, shown in
Table~\ref{tab:voice_profile_example}. Attribute-specific models refine demographics, emotion,
accent, and volume. Because the source corpora lack speaker labels, we cluster
WeSpeaker~\cite{wang2023wespeaker} embeddings within each session to obtain
pseudo speaker IDs. A separate utterance with the same pseudo ID is selected
as the speaker reference. This pipeline yields 20,046 bilingual examples;
model-specific annotation and clustering settings are provided in the
Appendix.

\paragraph{Annotation consistency.}
Three annotators assessed 1,162 bilingual utterances
(Table~\ref{tab:annotation_consistency}). Categorical attributes are evaluated
against human labels, whereas descriptive attributes are rated for
label--audio consistency on a 1--5 scale, with 5 indicating the closest match.

\begin{table}[h!]
\centering
\begin{minipage}[t]{0.55\textwidth}
    \vspace{0pt}
    \centering
    \caption{Example of a voice profile. The schema captures speaker, prosodic, affective, and contextual attributes.}
    \label{tab:voice_profile_example}
    \small
    \begin{tcolorbox}[
        colback=gray!8,
        colframe=black,
        boxrule=0.5pt,
        arc=1mm,
        top=3pt, bottom=3pt, left=6pt, right=6pt
    ]
    \scriptsize\ttfamily\raggedright
    \{ \\
    \hspace*{1.2em}"\textbf{gender}": "M", \\
    \hspace*{1.2em}"\textbf{age}": "young", \\
    \hspace*{1.2em}"\textbf{accent}": "mandarin", \\
    \hspace*{1.2em}"\textbf{emotion}": "Neutral", \\
    \hspace*{1.2em}"\textbf{emotion\_detail}": ["informative", "quiet pride"], \\
    \hspace*{1.2em}"\textbf{personality}": "Confident, resolute and...", \\
    \hspace*{1.2em}"\textbf{scene}": "A leader speaking at an assembly...", \\
    \hspace*{1.2em}"\textbf{speed}": "fast", \\
    \hspace*{1.2em}"\textbf{volume}": "normal", \\
    \hspace*{1.2em}"\textbf{texture}": "Bright, penetrating timbre..." \\
    \}
    \end{tcolorbox}
\end{minipage}\hfill
\begin{minipage}[t]{0.43\textwidth}
    \vspace{0pt}
    \centering
    \caption{Attribute-level consistency between human and model annotations. Categorical attributes are reported as accuracy (\%), and descriptive attributes on a 1--5 scale.}
    \vspace{5pt}
    \label{tab:annotation_consistency}
    \scriptsize
    \setlength{\tabcolsep}{1pt}
    \resizebox{\linewidth}{!}{%
    \begin{tabular}{lclc}
    \toprule
    \textbf{Attribute} & \textbf{Consistency $\uparrow$} &
    \textbf{Attribute} & \textbf{Consistency $\uparrow$} \\
    \midrule
    Gender      & 96.7 & Age         & 95.7 \\
    Accent      & 91.6 & Emotion     & 70.8 \\
    Speed       & 86.7 & Volume      & 72.5 \\
    \midrule
    Emo. Detail & 4.92 & Personality & 4.95 \\
    Texture     & 4.87 & Scene       & 4.79 \\
    \bottomrule
    \end{tabular}%
    }
\end{minipage}
\end{table}

\subsection{Joint Supervised Fine-Tuning}
\label{sec:phase1}
Joint SFT establishes instruction-conditioned generation and speech-to-profile recovery, providing a supervised initialization for reciprocal reinforcement learning, as illustrated in Figure~\ref{fig:cyclespeech_pipeline}(a). With teacher forcing, GEN
predicts $A^{*}$ from $(I,T,R)$, preserving the inference-time
instruction interface. At each joint step, GEN samples
$\widehat{A}\sim p_{\theta_G}(A\mid I,T,R;\phi)$, which the frozen token-to-wave
decoder converts into $\widehat{x}$. UND recovers the canonical JSON sequence
$Z(P^{*})$ from this online-generated waveform. Both branches use
length-normalized autoregressive token negative log-likelihoods:
\begin{equation}
\begin{aligned}
    \mathcal{L}_{\mathrm{Gen}}
    &= -\frac{1}{|A^{*}|}\sum_{t=1}^{|A^{*}|}
       \log p_{\theta_G}
       \left(a_t^{*}\mid I,T,R,a_{<t}^{*};\phi\right),\\
    \mathcal{L}_{\mathrm{Und}}
    &= -\frac{1}{|Z(P^{*})|}\sum_{m=1}^{|Z(P^{*})|}
       \log p_{\theta_U}
       \left(z_m^{*}\mid \operatorname{sg}[\widehat{x}],z_{<m}^{*};\phi\right).
\end{aligned}
\label{eq:joint-sft-token-losses}
\end{equation}
After the GEN-only warm-up, the joint objective is:
\begin{equation}
    \mathcal{L}_{\mathrm{SFT}}
    = \mathcal{L}_{\mathrm{Gen}}
    + \lambda_{\mathrm{und}}\mathcal{L}_{\mathrm{Und}},
    \qquad \lambda_{\mathrm{und}}=1.
    \label{eq:joint-sft-main}
\end{equation}
where $\operatorname{sg}$ stops gradients across the waveform interface, so
$\mathcal{L}_{\mathrm{Gen}}$ and $\mathcal{L}_{\mathrm{Und}}$ update only
$\theta_G$ and $\theta_U$, respectively. Because UND is trained on the current
generator's output distribution, unreliable early rollouts would provide an
unstable acoustic input. We therefore optimize GEN alone for the first 2,000
updates, allowing it to produce usable instruction-conditioned speech before
introducing online UND supervision. Both adapters are then jointly trained until step 10,000. This
design exposes UND to realistic generation artifacts while avoiding unstable
coupling at initialization.

\subsection{CycleGRPO Reciprocal Alignment}
CycleGRPO links two checks: whether generated speech realizes target
attributes, and whether inferred profiles preserve information that can
guide speaking-style reconstruction. Both adapters are initialized from
Joint SFT before alternating optimization begins. During each update, the
active adapter samples $M=6$ candidates while the other supplies
inference-only cycle feedback.

\subsubsection{Forward Loop: Synthesizer Alignment}
Naturalness alone does not establish that generated speech realizes the
requested vocal attributes. The forward loop assesses attribute mismatches
by recovering a profile from the waveform and comparing it with $P^{*}$.
GEN samples $M=6$ speech candidates from the instruction $I$, transcript
$T$, and speaker reference $R$, without conditioning on $P^{*}$. The current UND adapter then deterministically recovers a
voice profile from each waveform:
\begin{equation}
\begin{aligned}
    \widehat{A}_i
    &\sim \pi_{\theta_G}(\cdot\mid I,T,R),
    \qquad
    \widehat{x}_i
    = \operatorname{Dec}_{\mathrm{token2wav}}(\widehat{A}_i;R),\\
    \widehat{P}_i
    &= \operatorname{UND}_{\theta_U}(\widehat{x}_i),
    \qquad i=1,\ldots,M.
\end{aligned}
\label{eq:gen-cycle-main}
\end{equation}
Profile consistency supplies the attribute-control signal, while content
and quality terms penalize degradation in intelligibility and acoustic
quality. The resulting GEN reward is a weighted combination of these
complementary signals:
\begin{equation}
    R_i^{G}
    =0.70R_{\mathrm{prof}}(\widehat{P}_i,P^{*})
    +0.15\widetilde{R}_{\mathrm{con}}(\widehat{x}_i,T)
    +0.15R_{\mathrm{mos}}(\widehat{x}_i).
    \label{eq:gen-reward-main}
\end{equation}

\paragraph{Profile consistency.}
The canonical profile contains six categorical fields (gender, age, accent,
emotion, speed, and volume) and four free-text fields (emotion detail,
personality, scene, and vocal texture). List-valued descriptions are matched
before aggregation:
\begin{equation}
    R_{\mathrm{prof}}
    =\frac{1}{10}\left(
    \sum_{k\in\mathcal{K}_{\mathrm{cat}}}s_k
    +\sum_{k\in\mathcal{K}_{\mathrm{sem}}}s_k\right).
    \label{eq:profile-reward-main}
\end{equation}
Here, $\mathcal{K}_{\mathrm{cat}}$ and $\mathcal{K}_{\mathrm{sem}}$ index the
categorical and free-text fields, respectively. For field $k$, $s_k$ is an
exact-match indicator for categorical attributes or BGE-M3~\cite{bge-m3} cosine similarity
for free-text attributes. The factor $1/10$ averages the ten paralinguistic
fields equally, excluding language and transcript.

\paragraph{Content and quality guards.}
English and Chinese candidates are transcribed with Whisper-large-v3~\cite{radford2023robust} and
FireRedASR2S~\cite{xu2026fireredasr2s}, respectively. Let $\operatorname{ER}$ denote WER or CER. The
content reward saturates after high recognition accuracy:
\begin{equation}
\begin{aligned}
    R_{\mathrm{con}}^{\mathrm{raw}}
    &=\operatorname{clip}(1-\operatorname{ER},0,1),\\
    \widetilde{R}_{\mathrm{con}}
    &=\min\!\left(R_{\mathrm{con}}^{\mathrm{raw}}/0.95,1\right).
\end{aligned}
\label{eq:content-reward-main}
\end{equation}
This preserves a penalty for genuine content degradation without over-optimizing
small ASR differences. $R_{\mathrm{mos}}$ is the normalized UTMOS~\cite{saeki22c_interspeech}
score and discourages acoustically degraded solutions.

\subsubsection{Backward Loop: Comprehension Alignment}
Because profile matching alone does not establish controllability, the backward
loop tests whether inferred attributes support speaking-style reconstruction.
UND samples $M=6$ profiles from the gold waveform $x^{*}$ and scores them
against the fixed target $P^{*}$ to anchor training. A deterministic bridge
$\mathcal{B}$ converts each predicted profile and transcript $T$ into a
natural-language GEN input $\widetilde{I}_i$:
\begin{equation}
\begin{aligned}
    \widehat{P}_i
    &\sim \pi_{\theta_U}(\cdot\mid x^{*}),
    \qquad
    \widetilde{I}_i
    = \mathcal{B}(\widehat{P}_i,T),\\
    \widetilde{A}_i
    &\sim \pi_{\theta_G}(\cdot\mid\widetilde{I}_i,R),
    \qquad
    \widetilde{x}_i
    = \operatorname{Dec}_{\mathrm{t2w}}(\widetilde{A}_i;R).
\end{aligned}
\label{eq:und-cycle-main}
\end{equation}
GEN receives the predicted profile through $\widetilde{I}_i$.
The UND reward combines direct profile supervision with content and style feedback from the reconstructed speech:
\begin{equation}
    R_i^{U}
    =0.60R_{\mathrm{prof}}(\widehat{P}_i,P^{*})
    +0.20\widetilde{R}_{\mathrm{con}}(\widetilde{x}_i,T)
    +0.20R_{\mathrm{sty}}(\widetilde{x}_i,x^{*}).
    \label{eq:und-reward-main}
\end{equation}
The style term tests whether the inferred
attributes support faithful acoustic reconstruction. The content term guards
transcript fidelity during reconstruction, with $T$ supplied independently of
the inferred profile.

\paragraph{Style reconstruction.}
We use a frozen ParaMETA encoder~\cite{lou2026parameta} to measure the
paralinguistic speaking-style similarity between the GEN-reconstructed waveform
$\widetilde{x}_i$ and the ground-truth waveform $x^{*}$:
\begin{equation}
    R_{\mathrm{sty}}
    =\operatorname{Sim}_{\mathrm{ParaMETA}}
    (\widetilde{x}_i,x^{*}).
    \label{eq:style-reward-main}
\end{equation}
where $\operatorname{Sim}_{\mathrm{ParaMETA}}$ denotes similarity between the
ParaMETA speaking-style representations of the two waveforms. This reward is used
only in the backward loop. During these updates, UND is the active policy and
GEN remains inference-only, so the discrete profile and waveform interfaces
prevent gradients from crossing between adapters. Further reward normalization
details and bridge templates are provided in the Appendix.

\subsubsection{Alternating Optimization}
We construct the RL training pool using repeated Joint-SFT rollouts, discarding invalid, nearly solved, and reward-degenerate groups. The remaining prompts are stratified by difficulty, using mean reward as a proxy, and within-group reward variance. This yields 9,962 examples, comprising 4,976 GEN and 4,986 UND prompts.

On this pool, we alternate five GEN updates with five UND updates using the direction-specific rewards defined above. For each direction, a frozen copy of the corresponding Joint-SFT adapter serves as the reference policy for KL regularization.

Using $R_i^{G}$ for GEN or $R_i^{U}$ for UND, we compute a group-normalized
sequence advantage $A_i$ for each rollout. We optimize the active adapter
using the same clipped GRPO objective:
\begin{equation}
\begin{aligned}
    \mathcal{L}_{\mathrm{GRPO}}
    = \frac{1}{M}\sum_{i=1}^{M}\frac{1}{T_i}\sum_{t=1}^{T_i}
    \big[&-\min\big(r_{i,t}A_i,
    \operatorname{clip}(r_{i,t},1-\epsilon,1+\epsilon)A_i\big) +\beta\widehat{D}_{i,t}\big].
\end{aligned}
\label{eq:cyclegrpo-main}
\end{equation}
where $r_{i,t}$ is the policy ratio and $\widehat{D}_{i,t}$ is the sampled-action
KL penalty to the frozen Joint-SFT policy. The same $A_i$ is applied to
all $T_i$ generated tokens in rollout $i$. We use clipping threshold
$\epsilon=0.2$ and KL coefficient $\beta=0.04$ for both directions. Further implementation details are provided in the Appendix.

\section{Experiment}
\label{sec:results}
\subsection{Evaluation Setup and Metrics}
\label{sec:eval_setup}
Test speech and instructions are held out from the curated dataset. The GEN evaluation set contains
1,077 prompts, balanced across APS, DSD, and RP within each language. The UND evaluation set comprises 2,127 source utterances: 1,144 Chinese and 983 English. We compare
instruction-controlled generation and paralinguistic understanding baselines
(Tables~\ref{tab:main_results} and~\ref{tab:updated_metrics_evaluation}).

\paragraph{Automatic generation metrics.}
Content accuracy is measured by Chinese CER using FireRedASR2S~\cite{xu2026fireredasr2s} and English WER using Whisper-large-v3~\cite{radford2023robust}.
Speaker verification (Spk. Ver.) reports the percentage of generated/reference pairs with CAM++~\cite{wang2023cam++} speaker-embedding cosine similarity of at least 0.33.
Style Acc. measures exact-match accuracy over task-specific paralinguistic attributes including gender, age, accent, emotion, speed, and volume. Each applicable field receives a binary (0/1) exact-match score, averaged within each utterance.
Style SIM measures cosine similarity between temporally pooled FACodec~\cite{ju2024naturalspeech3} prosody representations of generated and ground-truth speech.
Instruction Match reports the percentage of positive instruction--speech consistency judgments from Gemini-2.5-pro, using the evaluation prompt provided by InstructTTSEval~\cite{huang2025instructttseval}.

\paragraph{Speech understanding metrics.}
PR measures the fraction of outputs parsed into profile objects.
For output $i$, the format score $F_i$ is the fraction of the 11 normalized output fields satisfying the schema; unparseable outputs receive $F_i=0$.
Schema adherence is reported as $\mathrm{SA}=100\,\mathrm{mean}_i(F_i)$ across all evaluated outputs.
Cat.Avg averages exact-match accuracy over gender, age, accent, emotion, speed, and volume.
Desc.Sim averages BGE-M3~\cite{bge-m3} cosine similarity over personality, scene, voice texture, and emotion detail.
TF is clipped transcript accuracy, $100\,\mathrm{clip}(1-\mathrm{CER},0,1)$.
Let $C_i$, $D_i$, and $T_i$ denote per-utterance Cat.Avg, Desc.Sim, and TF on a 0--100 scale.
OS averages $F_i(0.42C_i+0.28D_i+0.30T_i)$ over utterances.
Cat.Avg, Desc.Sim, TF, and OS are reported as utterance-macro percentages.

\section{Results and Analysis}
% Generation
\begin{table*}[h!]
\caption{Instruction-controlled TTS results. WER/CER and Style Acc. use the shared evaluation protocol in Section~\ref{sec:eval_setup}; brackets give signed 95\% confidence-bound offsets. Best and second-best results are \textbf{bolded} and \underline{underlined}.}
\label{tab:main_results}
\centering
\setlength{\tabcolsep}{2pt}
\resizebox{\textwidth}{!}{
\begin{tabular}{lccccc}
\toprule
{\textbf{Method / Model}}
& \shortstack{{\textbf{CER / WER}}{\textbf{(\%) $\downarrow$}}}
& \shortstack{{\textbf{Spk. Ver.}}{\textbf{(\%) $\uparrow$}}}
& \shortstack{{\textbf{Style Acc.}}{\textbf{(\%) $\uparrow$}}}
& \shortstack{{\textbf{Style SIM}}{$\uparrow$}}
& \shortstack{{\textbf{Instruction Match}}{\textbf{(\%) $\uparrow$}}}\\
\midrule

% =========================================================================
% ======================= PART I: CHINESE TEST SET =======================
% =========================================================================
\rowcolor[gray]{0.88}
\multicolumn{6}{l}{\textbf{Part I: Chinese Evaluation Benchmark}} \\
\addlinespace
% --- 基线 ---
GT Wav (upper reference) & 4.44 {\scriptsize [-0.99, +1.09]} & 99.00 {\scriptsize [-0.83, +0.67]} & 62.56 {\scriptsize [-2.59, +2.60]} & 1.00 {\scriptsize [0.00, 0.00]} & 81.50 {\scriptsize [-3.17, +3.00]} \\
VoxInstruct \cite{zhou2024voxinstruct} & 10.90 {\scriptsize [-1.61, +1.73]} & 80.17 {\scriptsize [-3.17, +3.17]} & 58.87 {\scriptsize [-2.37, +2.31]} & 0.87 {\scriptsize [-0.01, +0.01]} & 73.00 {\scriptsize [-3.50, +3.50]} \\
CosyVoice3 \cite{du2025cosyvoice3} & 2.76 {\scriptsize [-0.87, +1.17]} & \textbf{99.50} {\scriptsize [-0.67, +0.50]} & 56.28 {\scriptsize [-2.23, +2.21]} & 0.92 {\scriptsize [-0.01, +0.01]} & \textbf{82.17} {\scriptsize [-3.17, +3.00]} \\
OV-InstructTTS \cite{ren2026ovinstructtts} & 3.64 {\scriptsize [-0.76, +0.82]} & 99.00 {\scriptsize [-0.83, +0.67]} & 58.63 {\scriptsize [-2.21, +2.23]} & 0.90 {\scriptsize [-0.01, +0.01]} & 79.67 {\scriptsize [-3.33, +3.17]} \\
MiMo-Audio-Instruct \cite{zhang2025mimo} & 4.42 {\scriptsize [-0.77, +0.82]} & 81.83 {\scriptsize [-3.00, +3.00]} & 58.76 {\scriptsize [-2.13, +2.11]} & 0.92 {\scriptsize [-0.01, +0.01]} & \underline{80.33} {\scriptsize [-3.17, +3.17]} \\
\addlinespace
% --- Base 对比组 ---
\rowcolor[gray]{0.95}
\multicolumn{6}{l}{\textit{Comparison Group A: Optimization Against Base Architecture}} \\
Step-Audio-2-mini \cite{wu2025stepaudio2technicalreport} & \textbf{1.87} {\scriptsize [-0.53, +0.59]} & \underline{99.33} {\scriptsize [-0.67, +0.50]} & 58.07 {\scriptsize [-2.23, +2.23]} & \underline{0.93} {\scriptsize [-0.01, +0.01]} & 74.50 {\scriptsize [-3.50, +3.50]} \\
\textit{CycleSpeech (Phase 1: Joint SFT)} & 2.63 {\scriptsize [-0.63, +0.71]} & 98.83 {\scriptsize [-0.83, +0.83]} & \underline{60.20} {\scriptsize [-2.30, +2.31]} & \underline{0.93} {\scriptsize [-0.01, +0.01]} & 77.50 {\scriptsize [-3.33, +3.33]} \\
\textbf{CycleSpeech (Phase 2: CycleGRPO)} & \underline{2.33} {\scriptsize [-0.56, +0.65]} & \textbf{99.50} {\scriptsize [-0.67, +0.50]} & \textbf{60.42} {\scriptsize [-2.30, +2.29]} & \textbf{0.94} {\scriptsize [-0.01, +0.01]} & 79.00 {\scriptsize [-3.33, +3.33]} \\
\addlinespace
\midrule

% =========================================================================
% ======================= PART II: ENGLISH TEST SET =======================
% =========================================================================
\rowcolor[gray]{0.88}
\multicolumn{6}{l}{\textbf{Part II: English Evaluation Benchmark}} \\
\addlinespace
% --- 基线 ---
GT Wav (upper reference) & 6.47 {\scriptsize [-1.73, +2.00]} & 100.00 {\scriptsize [0.00, 0.00]} & 67.91 {\scriptsize [-2.93, +2.87]} & 1.00 {\scriptsize [0.00, 0.00]} & 82.39 {\scriptsize [-3.56, +3.35]} \\
VoxInstruct \cite{zhou2024voxinstruct} & 12.97 {\scriptsize [-2.30, +2.52]} & 82.81 {\scriptsize [-3.56, +3.35]} & 56.75 {\scriptsize [-2.56, +2.66]} & 0.89 {\scriptsize [-0.01, +0.01]} & 74.42 {\scriptsize [-3.56, +3.35]} \\
CosyVoice3 \cite{du2025cosyvoice3} & 3.42 {\scriptsize [-0.95, +1.07]} & \textbf{99.79} {\scriptsize [-0.42, +0.21]} & 58.15 {\scriptsize [-2.74, +2.69]} & \underline{0.93} {\scriptsize [-0.01, 0.00]} & 80.08 {\scriptsize [-3.56, +3.56]} \\
OV-InstructTTS \cite{ren2026ovinstructtts} & 11.02 {\scriptsize [-5.63, +10.12]} & 93.29 {\scriptsize [-2.31, +2.10]} & 55.44 {\scriptsize [-2.50, +2.53]} & 0.89 {\scriptsize [-0.02, +0.01]} & 78.41 {\scriptsize [-3.56, +3.56]} \\
MiMo-Audio-Instruct \cite{zhang2025mimo} & 7.00 {\scriptsize [-1.28, +1.42]} & 72.54 {\scriptsize [-4.19, +3.98]} & 56.37 {\scriptsize [-2.56, +2.54]} & 0.91 {\scriptsize [-0.01, +0.01]} & 77.57 {\scriptsize [-3.77, +3.77]} \\
\addlinespace
% --- Base 对比组 ---
\rowcolor[gray]{0.95}
\multicolumn{6}{l}{\textit{Comparison Group A: Optimization Against Base Architecture}} \\
Step-Audio-2-mini \cite{wu2025stepaudio2technicalreport} & \textbf{2.56} {\scriptsize [-0.82, +0.95]} & 96.86 {\scriptsize [-1.68, +1.46]} & 58.18 {\scriptsize [-2.74, +2.72]} & \textbf{0.94} {\scriptsize [-0.01, +0.01]} & 75.47 {\scriptsize [-3.77, +3.77]} \\
\textit{CycleSpeech (Phase 1: Joint SFT)} & \underline{2.74} {\scriptsize [-0.80, +0.96]} & \underline{98.95} {\scriptsize [-1.05, +0.84]} & \underline{60.35} {\scriptsize [-2.39, +2.31]} & 0.92 {\scriptsize [-0.01, +0.01]} & \underline{81.97} {\scriptsize [-3.56, +3.35]} \\
\textbf{CycleSpeech (Phase 2: CycleGRPO)} & 3.08 {\scriptsize [-0.92, +1.07]} & 98.53 {\scriptsize [-1.26, +1.05]} & \textbf{60.78} {\scriptsize [-2.41, +2.44]} & 0.92 {\scriptsize [-0.01, +0.01]} & \textbf{85.53} {\scriptsize [-3.14, +3.14]} \\
\addlinespace
\bottomrule
\end{tabular}
}
\end{table*}

\begin{table*}[t]
\caption{Speech paralinguistic understanding results. \textbf{PR}: parse rate; \textbf{SA}: schema adherence; \textbf{Cat.Avg}: six-field categorical accuracy; \textbf{Desc.Sim}: four-field BGE-M3 cosine similarity; \textbf{TF}: transcription fidelity; \textbf{OS}: overall score. Best and second-best results are \textbf{bolded} and \underline{underlined}.}
\label{tab:updated_metrics_evaluation}
\centering
\setlength{\tabcolsep}{6pt}
\resizebox{\linewidth}{!}{%
\begin{tabular}{lcccccc}
\toprule
\multirow{2}{*}{\textbf{Model}}
& \multicolumn{2}{c}{\textbf{Structure (\%) $\uparrow$}}
& \multicolumn{2}{c}{\textbf{Attributes (\%) $\uparrow$}}
& \multicolumn{1}{c}{\textbf{Transcript (\%) $\uparrow$}}
& \multicolumn{1}{c}{\textbf{Overall (\%) $\uparrow$}} \\
\cmidrule(lr){2-3}\cmidrule(lr){4-5}\cmidrule(lr){6-6}\cmidrule(lr){7-7}
& \textbf{PR} & \textbf{SA} & \textbf{Cat.Avg} & \textbf{Desc.Sim} & \textbf{TF} & \textbf{OS} \\
\midrule

% =========================================================================
% ======================= PART I: CHINESE TEST SET =======================
% =========================================================================
\rowcolor[gray]{0.88}
\multicolumn{7}{l}{\textbf{Part I: Chinese Evaluation Benchmark}} \\
\addlinespace
Qwen2.5-Omni & \textbf{100.0} & 93.1 & 58.52 {\scriptsize [-1.03, +1.03]} & 45.37 {\scriptsize [-0.45, +0.46]} & \underline{85.9} {\scriptsize [-0.82, +0.79]} & 58.76 {\scriptsize [-0.56, +0.54]} \\
MiMo-Audio-7B-Instruct & \underline{99.8} & \underline{99.6} & 58.10 {\scriptsize [-1.06, +1.02]} & 61.05 {\scriptsize [-0.36, +0.35]} & 83.0 {\scriptsize [-0.97, +0.93]} & 66.25 {\scriptsize [-0.60, +0.58]} \\
Kimi-Audio-7B-Instruct & 94.4 & 90.9 & 45.37 {\scriptsize [-1.43, +1.43]} & 53.54 {\scriptsize [-0.86, +0.84]} & 80.5 {\scriptsize [-1.49, +1.41]} & 56.26 {\scriptsize [-1.06, +1.06]} \\
\midrule
Step-Audio-2-mini & 99.0 & 91.9 & 55.46 {\scriptsize [-1.03, +1.03]} & 52.12 {\scriptsize [-0.60, +0.62]} & 77.8 {\scriptsize [-1.62, +1.59]} & 57.00 {\scriptsize [-0.76, +0.74]} \\
\textit{CycleSpeech (Phase 1: Joint SFT)} & \textbf{100.0} & \textbf{100.0} & \underline{73.56} {\scriptsize [-1.08, +1.09]} & \underline{71.90} {\scriptsize [-0.42, +0.40]} & \textbf{89.9} {\scriptsize [-0.75, +0.72]} & \underline{77.98} {\scriptsize [-0.57, +0.56]} \\
\textbf{CycleSpeech (Phase 2: CycleGRPO)} & \textbf{100.0} & \textbf{100.0} & \textbf{74.18} {\scriptsize [-1.06, +1.08]} & \textbf{72.30} {\scriptsize [-0.42, +0.42]} & \textbf{89.9} {\scriptsize [-0.75, +0.73]} & \textbf{78.36} {\scriptsize [-0.56, +0.55]} \\

% =========================================================================
% ======================= PART II: ENGLISH TEST SET =======================
% =========================================================================
\rowcolor[gray]{0.88}
\multicolumn{7}{l}{\textbf{Part II: English Evaluation Benchmark}} \\
\addlinespace
Qwen2.5-Omni & 98.9 & 86.6 & 55.41 {\scriptsize [-1.46, +1.44]} & 35.40 {\scriptsize [-1.14, +1.10]} & 83.9 {\scriptsize [-1.41, +1.33]} & 53.06 {\scriptsize [-1.09, +1.05]} \\
MiMo-Audio-7B-Instruct & 99.8 & \underline{97.3} & 65.26 {\scriptsize [-1.19, +1.17]} & 58.95 {\scriptsize [-0.60, +0.61]} & \underline{87.0} {\scriptsize [-1.00, +0.95]} & 68.35 {\scriptsize [-0.68, +0.69]} \\
Kimi-Audio-7B-Instruct & 99.0 & 91.2 & 58.44 {\scriptsize [-1.32, +1.34]} & 47.33 {\scriptsize [-0.77, +0.79]} & 84.2 {\scriptsize [-1.41, +1.38]} & 58.32 {\scriptsize [-0.82, +0.83]} \\
\midrule
Step-Audio-2-mini & \underline{99.9} & 91.7 & 62.83 {\scriptsize [-1.15, +1.15]} & 43.88 {\scriptsize [-0.51, +0.52]} & 86.9 {\scriptsize [-0.95, +0.93]} & 59.45 {\scriptsize [-0.57, +0.58]} \\
\textit{CycleSpeech (Phase 1: Joint SFT)} & \textbf{100.0} & \textbf{100.0} & \underline{70.19} {\scriptsize [-1.14, +1.14]} & \underline{71.36} {\scriptsize [-0.42, +0.41]} & \textbf{90.3} {\scriptsize [-0.91, +0.89]} & \underline{76.54} {\scriptsize [-0.59, +0.59]} \\
\textbf{CycleSpeech (Phase 2: CycleGRPO)} & \textbf{100.0} & \textbf{100.0} & \textbf{70.50} {\scriptsize [-1.14, +1.15]} & \textbf{71.94} {\scriptsize [-0.40, +0.40]} & \textbf{90.3} {\scriptsize [-0.91, +0.89]} & \textbf{76.84} {\scriptsize [-0.61, +0.60]} \\
\bottomrule
\end{tabular}%
}
\end{table*}

\subsection{Main Quantitative Results}
\paragraph{Speech Generation}
Table~\ref{tab:main_results} highlights complementary gains from the two training stages.
Joint SFT provides most of the Style Acc. improvement over the backbone.
CycleGRPO further strengthens instruction--speech consistency, improving English Instruction Match by 3.56 points over Joint SFT.
The resulting model leads the evaluated generation systems in Chinese Style Acc. and English Instruction Match, while maintaining strong speaker verification in both languages.
Relative to Joint SFT, CycleGRPO improves instruction following and
reduces Chinese CER, with a slight increase in English WER.

\paragraph{Speech Understanding and Profile Recovery}
Joint SFT substantially improves profile recovery over Step-Audio-2-mini, increasing Chinese/English OS by 20.98/17.09 points (Table~\ref{tab:updated_metrics_evaluation}).
CycleGRPO further refines both categorical and descriptive attributes.
CycleGRPO consequently achieves the highest Cat.Avg, Desc.Sim, and OS among the evaluated systems in both languages.
The higher OS with unchanged TF indicates that the additional gains come from paralinguistic attribute recovery, not transcription fidelity.
Together with the generation results, this supports improved instruction control alongside more accurate profile recovery.

\subsection{Out-of-Domain Generalization}
\label{sec:ood_generalization}
We evaluate cross-dataset generalization on the VccmDataset test set A~\cite{ji2024controlspeech}. As shown in Table~\ref{tab:ood_results}, \textit{CycleSpeech (Phase 1)} consistently outperforms Step-Audio-2-mini across all paralinguistic attributes. Building upon this, \textit{CycleSpeech (Phase 2)} achieves the highest accuracies in speed, volume, and emotion, accompanied by lower WER relative to Phase 1. Speaker similarity remains competitive across both phases. 

\subsection{Subjective Evaluation}

Five listeners evaluated 50 speech samples per system. Speech samples were presented in randomized order with system identities concealed. In the merged evaluation,
each system received 100 ratings in total, corresponding to two ratings per sample on average. The superscripts in Table~\ref{tab:subjective_mos} indicate the half-widths of the 95\% confidence intervals.

\textit{CycleSpeech} achieves the highest mean subjective ratings across instruction/style faithfulness, naturalness, and speaker similarity (Table~\ref{tab:subjective_mos}).
The gains over Joint SFT, including a 0.18-point increase in NMOS, indicate that improved perceived style adherence need not compromise naturalness or speaker similarity.
\begin{table}[t]
\centering
\begin{minipage}[t]{0.65\linewidth}
    \vspace{0pt}
    \centering
    \caption{Out-of-domain generation results across WER (\%), Speaker Similarity (SIM), and Attribute Accuracies (\%) for Pitch (PIT), Speed (SPD), Volume (VOL), and Emotion (EMO).}
    \label{tab:ood_results}
    \scriptsize
    \setlength{\tabcolsep}{2pt}
    \renewcommand{\arraystretch}{1.15}
    \resizebox{\linewidth}{!}{%
    \begin{tabular}{l cccccc}
    \toprule
    \textbf{Model} & \textbf{WER$\downarrow$} & \textbf{SIM$\uparrow$} & \textbf{PIT$\uparrow$} & \textbf{SPD$\uparrow$} & \textbf{VOL$\uparrow$} & \textbf{EMO$\uparrow$} \\
    \midrule
    GT Codec                       & 3.47$^{\pm 0.35}$ & .970$^{\pm .002}$ & 90.2$^{\pm 1.5}$ & 88.7$^{\pm 1.6}$ & 89.7$^{\pm 1.5}$ & 72.6$^{\pm 6.3}$ \\
    Step-Audio-2-mini              & \textbf{1.93}$^{\pm 0.40}$ & \textbf{.860}$^{\pm .005}$ & 69.7$^{\pm 2.4}$ & 65.6$^{\pm 2.4}$ & 63.2$^{\pm 2.4}$ & 33.7$^{\pm 6.8}$ \\
    \midrule
    \textit{CycleSpeech (Phase1)} & 2.66$^{\pm 0.40}$ & \underline{.852}$^{\pm .005}$ & \textbf{73.1}$^{\pm 2.2}$ & \underline{67.1}$^{\pm 2.4}$ & \underline{65.6}$^{\pm 2.4}$ & \underline{37.4}$^{\pm 6.8}$ \\
    \textbf{CycleSpeech (Phase2)} & \underline{2.50}$^{\pm 0.40}$ & .851$^{\pm .006}$ & \underline{72.5}$^{\pm 2.3}$ & \textbf{68.3}$^{\pm 2.4}$ & \textbf{66.0}$^{\pm 2.4}$ & \textbf{40.0}$^{\pm 7.1}$ \\
    \bottomrule
    \end{tabular}%
    }
\end{minipage}\hfill
\begin{minipage}[t]{0.33\linewidth}
      \vspace{0pt}
      \centering
      \caption[Subjective mean opinion scores.]{\raggedright
      Subjective MOS for instruction/style faithfulness (FMOS),
      naturalness (NMOS), and speaker similarity (SMOS).
      \par}
      \label{tab:subjective_mos}
      \scriptsize
      \setlength{\tabcolsep}{3pt}
      \resizebox{\linewidth}{!}{%
      \begin{tabular}{lccc}
      \toprule
      \textbf{Model} & \textbf{FMOS} $\uparrow$ &
      \textbf{NMOS} $\uparrow$ & \textbf{SMOS} $\uparrow$ \\
      \midrule
      CosyVoice3 &
  3.82$^{\pm .24}$ & 3.91$^{\pm .21}$ & 3.71$^{\pm .19}$ \\

  Step-Audio-2-mini &
  3.61$^{\pm .20}$ & 3.88$^{\pm .17}$ & 3.53$^{\pm .18}$ \\
  \midrule
  \textit{CycleSpeech (phase1)} &
  3.99$^{\pm .21}$ & 4.07$^{\pm .21}$ & 3.90$^{\pm .22}$ \\
  \textbf{CycleSpeech (phase2)} &
  \textbf{4.15}$^{\pm .21}$ &
  \textbf{4.25}$^{\pm .17}$ &
  \textbf{4.00}$^{\pm .20}$ \\
  \bottomrule
      \end{tabular}%
      }
  \end{minipage}
  \end{table}

\subsection{Efficacy of CycleGRPO Closed-Loop Alignment}

\begin{table}[t]
\caption{Effect of the alternating update interval in CycleGRPO. Alt-$k$ switches the optimized branch every $k$ updates.}
\label{tab:alternation_interval_ablation}
\centering
\small
\setlength{\tabcolsep}{2.5pt}
\resizebox{\columnwidth}{!}{%
\begin{tabular}{lccccc}
\toprule
& \multicolumn{2}{c}{\textbf{GEN}} & \multicolumn{3}{c}{\textbf{UND}} \\
\cmidrule(lr){2-3} \cmidrule(lr){4-6}
\textbf{Variant}
& \textbf{WER/CER (\%)} $\downarrow$
& \textbf{Style Acc.} $\uparrow$
& \textbf{Cat. Avg.} $\uparrow$
& \textbf{Desc.Sim} $\uparrow$
& \textbf{OS} $\uparrow$ \\
\midrule
\multicolumn{6}{c}{\textit{\textbf{Chinese Test Set}}} \\
\midrule
CycleGRPO (Alt-1)
& \underline{2.47} {\scriptsize [-0.60, +0.67]}
& 58.82 {\scriptsize [-2.30, +2.28]}
& 74.08 {\scriptsize [-1.06, +1.08]}
& \textbf{72.37} {\scriptsize [-0.41, +0.40]}
& \textbf{78.36} {\scriptsize [-0.56, +0.55]} \\
CycleGRPO (Alt-10)
& 2.53 {\scriptsize [-0.60, +0.66]}
& \underline{59.69} {\scriptsize [-2.23, +2.21]}
& \underline{74.14} {\scriptsize [-1.08, +1.05]}
& 72.29 {\scriptsize [-0.42, +0.40]}
& \underline{78.31} {\scriptsize [-0.56, +0.55]} \\
\textbf{CycleGRPO (Alt-5, Ours)}
& \textbf{2.33} {\scriptsize [-0.56, +0.65]}
& \textbf{60.42} {\scriptsize [-2.30, +2.29]}
& \textbf{74.18} {\scriptsize [-1.06, +1.08]}
& \underline{72.30} {\scriptsize [-0.42, +0.42]}
& \textbf{78.36} {\scriptsize [-0.56, +0.55]} \\
\midrule
\multicolumn{6}{c}{\textit{\textbf{English Test Set}}} \\
\midrule
CycleGRPO (Alt-1)
& \textbf{2.85} {\scriptsize [-0.86, +1.01]}
& \textbf{61.21} {\scriptsize [-2.50, +2.42]}
& 70.30 {\scriptsize [-1.14, +1.14]}
& \textbf{71.95} {\scriptsize [-0.41, +0.40]}
& \underline{76.74} {\scriptsize [-0.59, +0.58]} \\
CycleGRPO (Alt-10)
& 3.21 {\scriptsize [-0.92, +1.07]}
& 59.55 {\scriptsize [-2.66, +2.60]}
& \textbf{70.65} {\scriptsize [-1.14, +1.12]}
& 71.79 {\scriptsize [-0.41, +0.40]}
& \textbf{76.84} {\scriptsize [-0.59, +0.59]} \\
\textbf{CycleGRPO (Alt-5, Ours)}
& \underline{3.08} {\scriptsize [-0.92, +1.07]}
& \underline{60.78} {\scriptsize [-2.41, +2.44]}
& \underline{70.50} {\scriptsize [-1.14, +1.15]}
& \underline{71.94} {\scriptsize [-0.40, +0.40]}
& \textbf{76.84} {\scriptsize [-0.61, +0.60]} \\
\bottomrule
\end{tabular}%
}
\end{table}

\paragraph{Alternating Optimization Dynamics.}
Alt-1/5/10 share reward weights, sampling hyperparameters, and the training pool. Overall profile recovery varies little across intervals, whereas generation is more sensitive to the switching frequency (Table~\ref{tab:alternation_interval_ablation}).
Alt-5 delivers the strongest Chinese generation performance and achieves the joint-highest OS in both languages, offering the best overall balance across languages and tasks.

\begin{table}[h!]
\caption{Effects of reciprocal cycle feedback, evolving feedback models, and bidirectional adapter updates. w/o Cycle removes both feedback loops but retains alternating updates ($k=5$). w/ Frozen FB retains both cycles and alternating updates, with feedback from frozen Joint-SFT counterparts. w/o UND/GEN Updates trains only GEN/UND, respectively, with the opposite adapter frozen at Joint-SFT.}
\label{tab:cycle_direction_ablation}
\centering
\small
\setlength{\tabcolsep}{2.5pt}
\resizebox{\columnwidth}{!}{%
\begin{tabular}{lccccc}
\toprule
& \multicolumn{2}{c}{\textbf{GEN}} & \multicolumn{3}{c}{\textbf{UND}} \\
\cmidrule(lr){2-3} \cmidrule(lr){4-6}
\textbf{Variant}
& \textbf{WER/CER (\%)} $\downarrow$
& \textbf{Style Acc.} $\uparrow$
& \textbf{Cat. Avg.} $\uparrow$
& \textbf{Desc.Sim} $\uparrow$
& \textbf{OS} $\uparrow$ \\
\midrule
\multicolumn{6}{c}{\textit{\textbf{Chinese Test Set}}} \\
\midrule
w/o Cycle
& \textbf{2.29} {\scriptsize [-0.56, +0.61]}
& 58.92 {\scriptsize [-2.26, +2.18]}
& 74.01 {\scriptsize [-1.09, +1.09]}
& \textbf{72.45} {\scriptsize [-0.41, +0.40]}
& \underline{78.35} {\scriptsize [-0.56, +0.55]} \\
w/ Frozen FB
& 2.74 {\scriptsize [-0.70, +0.79]}
& 59.66 {\scriptsize [-2.24, +2.24]}
& \underline{74.08} {\scriptsize [-1.08, +1.08]}
& \underline{72.36} {\scriptsize [-0.40, +0.40]}
& 78.34 {\scriptsize [-0.56, +0.55]} \\
w/o UND Updates
& 2.46 {\scriptsize [-0.58, +0.65]}
& 59.69 {\scriptsize [-2.21, +2.17]}
& 73.56 {\scriptsize [-1.08, +1.09]}
& 71.90 {\scriptsize [-0.42, +0.40]}
& 77.98 {\scriptsize [-0.57, +0.56]} \\
w/o GEN Updates
& 2.63 {\scriptsize [-0.63, +0.71]}
& \underline{60.20} {\scriptsize [-2.30, +2.31]}
& 73.97 {\scriptsize [-1.06, +1.08]}
& 72.26 {\scriptsize [-0.41, +0.40]}
& 78.25 {\scriptsize [-0.56, +0.56]} \\
\textbf{CycleGRPO (Ours)}
& \underline{2.33} {\scriptsize [-0.56, +0.65]}
& \textbf{60.42} {\scriptsize [-2.30, +2.29]}
& \textbf{74.18} {\scriptsize [-1.06, +1.08]}
& 72.30 {\scriptsize [-0.42, +0.42]}
& \textbf{78.36} {\scriptsize [-0.56, +0.55]} \\
\midrule
\multicolumn{6}{c}{\textit{\textbf{English Test Set}}} \\
\midrule
w/o Cycle
& \underline{2.89} {\scriptsize [-0.93, +1.11]}
& 58.96 {\scriptsize [-2.53, +2.42]}
& 70.19 {\scriptsize [-1.14, +1.14]}
& 71.91 {\scriptsize [-0.41, +0.40]}
& 76.69 {\scriptsize [-0.60, +0.59]} \\
w/ Frozen FB
& 2.91 {\scriptsize [-0.89, +1.06]}
& \underline{60.68} {\scriptsize [-2.55, +2.46]}
& \underline{70.26} {\scriptsize [-1.15, +1.12]}
& \underline{71.92} {\scriptsize [-0.43, +0.42]}
& 76.72 {\scriptsize [-0.61, +0.60]} \\
w/o UND Updates
& 2.91 {\scriptsize [-0.87, +1.03]}
& 59.53 {\scriptsize [-2.46, +2.46]}
& 70.19 {\scriptsize [-1.14, +1.14]}
& 71.36 {\scriptsize [-0.42, +0.41]}
& 76.54 {\scriptsize [-0.59, +0.59]} \\
w/o GEN Updates
& \textbf{2.74} {\scriptsize [-0.80, +0.96]}
& 60.35 {\scriptsize [-2.39, +2.31]}
& \textbf{70.50} {\scriptsize [-1.10, +1.12]}
& 71.88 {\scriptsize [-0.41, +0.41]}
& \underline{76.81} {\scriptsize [-0.59, +0.58]} \\
\textbf{CycleGRPO (Ours)}
& 3.08 {\scriptsize [-0.92, +1.07]}
& \textbf{60.78} {\scriptsize [-2.41, +2.44]}
& \textbf{70.50} {\scriptsize [-1.14, +1.15]}
& \textbf{71.94} {\scriptsize [-0.40, +0.40]}
& \textbf{76.84} {\scriptsize [-0.61, +0.60]} \\
\bottomrule
\end{tabular}%
}
\end{table}

\paragraph{Impact of Reciprocal Cycle Rewards.}
The \textit{w/o Cycle} variant retains alternating GRPO updates for both adapters but disables both feedback loops.
Generated speech is not verified by UND, and predicted profiles are not used by GEN for reconstruction.
GEN uses equally weighted ASR-based content and UTMOS quality rewards; UND uses only ground-truth profile matching.
\textit{w/ Frozen FB} retains both cycles, alternating updates of both adapters, and CycleGRPO's training budget, but obtains feedback from frozen Joint-SFT counterparts. The \textit{w/o UND Updates} variant updates only GEN with UND frozen at Joint-SFT; \textit{w/o GEN Updates} reverses these roles.
Each retains the active adapter's full reward objective, weights, and KL regularization.

CycleSpeech leads Style Acc. in both languages, attains the highest Chinese Cat.Avg, and ties the highest English Cat.Avg (Table~\ref{tab:cycle_direction_ablation}).
Against \textit{w/o Cycle}, Style Acc. improves by 1.50/1.82 points on Chinese/English, supporting the contribution of cycle feedback to attribute control.
Both single-direction controls remain below CycleSpeech in Style Acc.; w/o GEN Updates ties its English Cat.Avg.
Comparison with \textit{w/ Frozen FB} more directly tests the value of evolving feedback. Frozen counterparts(\textit{w/ Frozen FB}) already improve Style Acc. over \textit{w/o Cycle}, but updating them yields further gains in both Style Acc. and Cat.Avg.

\section{Conclusion}
We presented CycleSpeech, which unifies instruction-controlled speech synthesis and paralinguistic understanding through a shared, recoverable voice profile. The forward cycle verifies intended attributes in generated speech, while the backward cycle tests whether inferred profiles support speaking-style reconstruction. Building on Joint SFT, CycleGRPO converts these checks into reciprocal training feedback without human preference annotations or a separately trained preference reward model. Our aligned bilingual corpus supports both understanding supervision and generation verification. Bilingual evaluations and controlled ablations show improved instruction following and categorical profile recovery, alongside strong subjective naturalness and speaker similarity. Together, these findings support reciprocal verification through recoverable attributes as a practical approach to controllable speech generation.

\section*{Acknowledgments}
We sincerely thank Zhuo Chen, Xiaobin Zhuang, Dongya Jia, and Yuanzhe Chen from the ByteDance Seed team for their valuable guidance and constructive feedback. Their suggestions helped us sharpen the research focus and strengthen the experimental design. We greatly appreciate their time, expertise, and thoughtful engagement with this work.

\bibliographystyle{plainnat}
\bibliography{references}

\clearpage
\appendix
% Supplementary content included by main_public.tex.
\makeatletter
\setlength{\@fptop}{0pt}
\setlength{\@fpsep}{14pt}
\setlength{\@fpbot}{0pt plus 1fil}
\makeatother

\setcounter{equation}{0}
\setcounter{table}{0}
\setcounter{figure}{0}
\renewcommand{\theequation}{S\arabic{equation}}
\renewcommand{\thetable}{S\arabic{table}}
\renewcommand{\thefigure}{S\arabic{figure}}

\section{Paradigm Comparison}
\label{sec:supp_paradigm_comparison}
Table~\ref{tab:paradigm_comparison} compares the functional roles of generation,
understanding, shared targets, and reciprocal optimization across paradigms.

\begin{table}[htbp]
\centering
    \caption{Paradigm comparison across RL-aligned instruction-controlled generation and paralinguistic evaluation or understanding. \textit{Shared Target} denotes a semantic target shared by generation and understanding; \textit{Cycle Opt.} denotes reciprocal optimization between them.}
    \label{tab:paradigm_comparison}
    \small
    \setlength{\tabcolsep}{5pt}
    \renewcommand{\arraystretch}{1.05}
    \begin{tabularx}{\linewidth}{
        @{} >{\raggedright\arraybackslash}p{0.27\linewidth}
        c c c >{\columncolor[gray]{0.97}}c
        >{\raggedright\arraybackslash}X @{} }
    \toprule
    \textbf{Method} & \textbf{Gen.} & \textbf{Und.} &
    \shortstack{\textbf{Shared}\\\textbf{Target}} &
    \shortstack{\textbf{Cycle}\\\textbf{Opt.}} & \textbf{Reward} \\
    \midrule
    \rowcolor[gray]{0.93}
    \multicolumn{6}{@{}l}{\textit{RL-aligned generation}} \\
    CosyVoice~3 & \yes & \no & \no & \no & ASR/SER scores \\
    Fish Audio~S2 & \yes & \no & \no & \no & ASR, quality, speaker \\
    FlexiVoice & \yes & \no & \no & \no & Emotion and timbre preference \\
    \midrule
    \rowcolor[gray]{0.93}
    \multicolumn{6}{@{}l}{\textit{Evaluation and understanding}} \\
    InstructTTS\-Eval & \no & \yes & \no & \no & External audio-LLM \\
    ALLM Judges & \no & \yes & \no & \no & Instruction and style rating \\
    StyleCap & \no & \yes & \no & \no & Free-form caption \\
    ParaSpeech\-CLAP & \no & \yes & \no & \no & Speech--caption similarity \\
    \midrule
    \rowcolor[gray]{0.93}
    \multicolumn{6}{@{}l}{\textit{Joint generation and understanding}} \\
    UniStyle & \yes & \yes & \yes & \no & Free-form caption \\
    \textbf{CycleSpeech} & \textbf{\yes} & \textbf{\yes} &
    \textbf{\yes} & \textbf{\yes} & \textbf{Structured profile} \\
    \bottomrule
    \end{tabularx}
\end{table}

\section{Training Data Construction}
\label{sec:supp_data_construction}

\begin{figure}[h!]
    \centering
    \includegraphics[width=\linewidth]{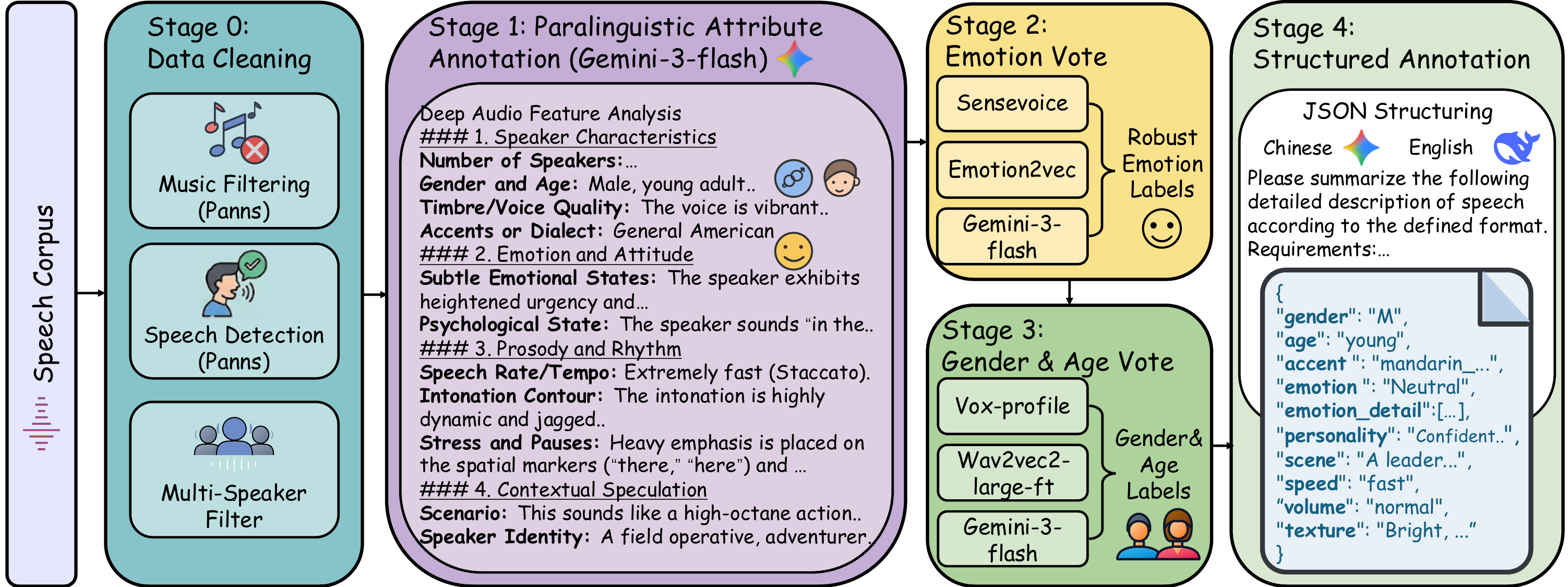}
    \caption{Data annotation pipeline.}
    \label{fig:data_construction_pipeline}
\end{figure}

A paired training instance is
$\mathcal{D}_i=(I_i,T_i,R_i,x_i^{*},A_i^{*},P_i^{*})$, containing an
instruction, transcript, speaker reference, target waveform, target speech
tokens, and reference voice profile, respectively. The shared backbone has
frozen parameters $\phi$; $\theta_G$ and $\theta_U$ parameterize the GEN and
UND adapters. Figure~\ref{fig:data_construction_pipeline} summarizes the
annotation pipeline.

\subsection{Data Sources, Filtering, and Composition}
\label{sec:supp_data_filtering}
Chinese speech is drawn from WenetSpeech, and English speech from an internal
collection of film and television recordings. Full-length English recordings
are first segmented using subtitle timestamps, and the corresponding subtitle
text serves as the transcript. 

Audio is converted to 16-kHz mono. We use PANNs Cnn14 to filter audio clips, retaining only those with a predicted speech probability of at least 0.5 and a predicted music probability below 0.5. We retain only single-speaker audio clips,
excluding those containing multiple speakers.

The paired SFT set contains 20,046 source utterances. The GRPO pool contains 9,962 direction-specific
records, comprising 4,976 GEN and 4,986 UND prompts. Table~\ref{tab:supp_data_composition}
summarizes their instruction counts and session-local pseudo speaker IDs.
The observed duration ranges are 1.00--9.98\,s for SFT and 1.00--10.152\,s for
GRPO.

\begin{table}[htbp]
\caption{Training-set composition.}
\label{tab:supp_data_composition}
\centering
\small
\setlength{\tabcolsep}{7pt}
\begin{tabular}{lrrrrr}
\toprule
\textbf{Stage} & \textbf{APS} & \textbf{DSD} & \textbf{RP} &
\textbf{Total} & \textbf{Pseudo IDs} \\
\midrule
Joint SFT & 6,047 & 8,266 & 5,733 & 20,046 & 17,043 \\
CycleGRPO & 3,791 & 3,571 & 2,600 & 9,962 & 7,996 \\
\bottomrule
\end{tabular}
\end{table}

\subsection{Profile Annotation and Canonical Fields}
\label{sec:supp_profile_annotation}
Gemini-3-flash produces an auditory report covering speaker characteristics,
affect, prosody, vocal texture, speaking behavior, and inferred communicative
context. DeepSeek-v3 and Gemini-3-flash convert the Chinese and English reports, respectively, into structured voice profiles following a shared JSON schema.

\paragraph{Attribute refinement.}
Gender and age labels are refined by majority voting among predictions from
Gemini-3-flash, WavLM-large-age-sex (Vox-Profile) and
wav2vec2-large-robust-24-ft-age-gender. Emotion predictions from Gemini-3-flash, SenseVoice, and
Emotion2Vec are mapped to seven core categories plus an "Other" category.
Majority voting combines these predictions, with ties resolved using the Gemini-3-flash annotation.
FireRedLID supplies Chinese accent or dialect labels, whereas English accent labels are taken directly from Gemini-3-flash annotations. The canonical label sets
are listed in Table~\ref{tab:supp_profile_labels}.

\begin{table}[h!]
\caption{Canonical categorical label sets.}
\label{tab:supp_profile_labels}
\centering
\footnotesize
\renewcommand{\arraystretch}{1.12}
\begin{tabularx}{\linewidth}{@{}p{0.17\linewidth}X@{}}
\toprule
\textbf{Field} & \textbf{Canonical values} \\
\midrule
Gender & \texttt{M}, \texttt{F}, \texttt{unk} \\
Age & \texttt{child}, \texttt{teen}, \texttt{young}, \texttt{middle}, \texttt{senior}, \texttt{unknown} \\
Speed & \texttt{slow}, \texttt{medium\_slow}, \texttt{medium}, \texttt{medium\_fast}, \texttt{fast}, \texttt{unknown} \\
Volume & \texttt{quiet}, \texttt{normal}, \texttt{loud}, \texttt{unknown} \\
Emotion & \texttt{Neutral}, \texttt{Happy}, \texttt{Sad}, \texttt{Angry}, \texttt{Surprised}, \texttt{Fearful}, \texttt{Disgusted}, \texttt{Other} \\
English accent & \texttt{english\_general\_american}, \texttt{english\_british}, \texttt{english\_australian}, \texttt{english\_indian}, \texttt{english\_mid\_atlantic}, \texttt{english\_other}, \texttt{unknown} \\
Chinese accent & \texttt{mandarin}; \texttt{regional\_} followed by \texttt{northern}, \texttt{beijing}, \texttt{northeast}, \texttt{taiwan}, \texttt{cantonese}, \texttt{southwest}, \texttt{hongkong}, \texttt{shaanxi}, \texttt{hunan}, \texttt{henan}, \texttt{sichuan}, \texttt{shandong}, \texttt{wu}, \texttt{min}, \texttt{other}, or \texttt{shanghai}; \texttt{unknown} \\
\bottomrule
\end{tabularx}
\end{table}

\paragraph{Profile fields and annotation metadata.}
The profile reward scores six categorical fields (gender, age, accent,
emotion, speed, and volume) and four descriptive fields (emotion detail,
personality, scene, and texture). The \texttt{emotion\_detail} field is an
array of one to three strings; the other descriptive fields are free text.
SA evaluates 11 normalized fields: these ten paralinguistic fields and the
transcript field. TF evaluates the predicted transcript.

\subsection{Speaker Annotation and Reference Construction}
\label{sec:supp_speaker_annotation}
WeSpeaker extracts embeddings using a CnCeleb-trained ResNet34 for Chinese
and a VoxCeleb-trained ResNet34 for English. Utterances of at most 15\,s use
one utterance-level embedding. Longer recordings use sliding-window
embeddings with spectral clustering. Embeddings are then clustered within
their source session using cosine-similarity thresholds of 0.55 for Chinese
and 0.65 for English.

For each SFT target, reference selection first searches for another utterance
with the same pseudo ID in the selected dataset. If none exists, it searches
the full speaker mapping. The first eligible candidate is selected, excluding
the target itself; no random, medoid, or quality-based ranking is applied.
Targets without an independent same-speaker reference are excluded from the
paired SFT set. All 20,046 retained SFT pairs have a nonempty reference that
differs from the target and shares its pseudo speaker ID.

\subsection{Instruction Construction and Annotation Validation}
\label{sec:supp_instruction_construction}
APS, DSD, and RP instructions are assembled through deterministic templates
from existing annotations; this assembly step does not invoke an LLM.
The inputs include the annotated acoustic attributes, emotion summaries,
prosodic details, personality descriptions, and role-play descriptions.
APS combines gender, age, pitch or timbre, speaking rate, volume, and accent.
DSD organizes timbre, rhythm, emotion, personality, and global prosody into a
natural-language style directive. RP extracts a short scenario or role from
the role-play description, removing dialogue and overly specific performance
details. Representative bilingual examples appear in
Table~\ref{tab:instruction_examples}.

\begin{table}[t]
    \caption{Examples of Chinese and English
    instructions from the APS, DSD, and RP categories. APS specifies acoustic
    attributes, DSD describes speaking style, and RP provides a communicative
    scenario.}
    \label{tab:instruction_examples}
    \centering
    \footnotesize
    \setlength{\tabcolsep}{4pt}
    \renewcommand{\arraystretch}{1.08}
    \begin{CJK*}{UTF8}{gbsn}
    \begin{tabularx}{\linewidth}{@{}>{\bfseries}p{0.12\linewidth}>{\bfseries}p{0.09\linewidth}X@{}}
    \toprule
    \textbf{Source} & \textbf{Type} & \textbf{Instruction} \\
    \midrule
    Chinese & APS & 女性，青壮年（20-30岁）；音色与音高偏高、清脆、明亮、金属质感；语速偏快，音量中等偏强；标准普通话。 \\
    & DSD & 以中等音高为主，音色上体现柔和、纤细、湿润，音量极低，展现出这段语音带有明显的焦虑和紧迫感与秘密感交织的情感层次，性格层面显得紧张、脆弱、警示者，音频中，说话人语速较快，带有宿命般的强调。 \\
    & RP & 在古装剧或宫廷题材场景中，一位青年女性角色在私密室内与心腹密探对话，听到的惊人消息后，她眉头微蹙、眼神凝滞，缓慢而沉重地复述这一事实，声音中充满震惊与疑虑，试图确认核实。 \\
    \midrule
    English & APS & gender: Female; age: early to mid-twenties; timbre / pitch: Bright and resonant, with a gentle terminal fall, Crystalline, smooth, slightly breathy, with a subtle hint of nasality; speech rate: Moderate and measured; volume: Normal conversational level; accent annotation: Standard General American accent. \\
    & DSD & Use a bright delivery in a General American accent, in a mid register, with measured pacing, at steady volume. Keep the tone grounded and matter-of-fact, with a relaxed presence. \\
    & RP & a young female content creator in her early twenties, sitting comfortably in her room, recording a casual vlog update for her followers. \\
    \bottomrule
    \end{tabularx}
    \end{CJK*}
\end{table}

\section{Additional Training and Reward Details}
\label{sec:supp_method_details}

\subsection{Joint SFT Sampling and Warm-Start Schedule}
\label{sec:supp_joint_sft}
The token-level negative log-likelihoods $\mathcal{L}_{\mathrm{Gen}}$ and
$\mathcal{L}_{\mathrm{Und}}$ are defined in Section~3.3 of the main paper. $\mathcal{L}_{\mathrm{Und}}$ denotes supervision
on online-generated speech, with $\operatorname{sg}[\widehat{x}]$ as input.
The frozen token-to-wave decoder converts a GEN rollout into a waveform:
\begin{equation}
    \widehat{A}\sim p_{\theta_G}(A\mid I,T,R;\phi),\qquad
    \widehat{x}=\operatorname{Dec}_{\mathrm{t2w}}(\widehat{A};R).
    \label{eq:supp_phase1_online_synth}
\end{equation}
The rollout uses temperature 0.7, top-$p=1.0$, top-$k=50$, and repetition
penalty 1.05. Its token budget depends on transcript length and is capped at
2,048 new tokens. UND uses the official Step-Audio-2 chat template.
GEN first synthesizes $\widehat{x}$ without gradient tracking.
UND is trained on the detached waveform
$\operatorname{sg}[\widehat{x}]$, so its loss updates only the UND
adapter and does not backpropagate into GEN.

Let $S_G$ and $S_U$ denote GEN-only warm-up updates and optional UND-only
warm-up updates on gold speech. At training step $s$, the schedule is
\begin{equation}
\mathcal{L}_{\mathrm{SFT}}^{(s)}=
\begin{cases}
    \mathcal{L}_{\mathrm{Gen}},
    &0\leq s<S_G,\\[1mm]
    \lambda_{\mathrm{und}}\mathcal{L}_{\mathrm{Und}}^{\mathrm{gold}},
    &S_G\leq s<S_G+S_U,\\[1mm]
    \mathcal{L}_{\mathrm{Gen}}+
    \lambda_{\mathrm{und}}\mathcal{L}_{\mathrm{Und}},
    &s\geq S_G+S_U.
\end{cases}
\label{eq:supp_phase1_schedule}
\end{equation}
Here, $\mathcal{L}_{\mathrm{Und}}^{\mathrm{gold}}$ uses the ground-truth
waveform $x^*$, rather than the detached GEN waveform
$\operatorname{sg}[\widehat{x}]$, as input to UND.
The selected setting is $S_G=2{,}000$, $S_U=0$, and
$\lambda_{\mathrm{und}}=1$.

Both adapters use LoRA rank 8 and dropout 0.05. GEN targets all linear layers
with LoRA scaling parameter $\alpha=32$; UND targets self-attention
$q/k/v/o$ projections with $\alpha=16$. AdamW uses learning rate
$2\times10^{-5}$, weight decay 0.01, gradient clipping at 1.0, a 2,000-step
linear learning-rate warm-up, and cosine decay. Learning-rate warm-up and the
GEN-only warm-up are distinct schedules, although both last 2,000 steps.
Training uses BF16 on four A800 GPUs with batch size 1 per worker and global
batch size 4. Maximum sequence lengths are 6,144 for GEN and 2,048 for UND.
The selected 10k-step Joint-SFT checkpoint initializes both CycleGRPO
adapters from their respective weights.

\subsection{CycleGRPO Optimization}
During online training, we alternately update GEN and UND, switching adapters every $k=5$ updates. Within each block, only the active adapter is optimized; the other provides cycle feedback using its latest weights without parameter updates. When providing feedback to GEN, UND decodes profiles deterministically. During its own update blocks, UND instead samples candidate profiles for GRPO. Separate Joint-SFT copies, $\pi_G^{\mathrm{ref}}$ and $\pi_U^{\mathrm{ref}}$, remain frozen throughout training and are used only for KL regularization.

Each of four data-parallel workers samples $M=6$ candidates for one prompt,
giving 24 candidates per synchronized global step. For the active direction,
$R_i$ denotes $R_i^G$ or $R_i^U$ as defined in the main paper. Statistics are
computed within each six-candidate prompt group, not across unrelated prompts:
\begin{equation}
\begin{gathered}
    \bar{R}=\frac{1}{M}\sum_{i=1}^{M}R_i,\qquad
    \sigma_R=\sqrt{\frac{1}{M}\sum_{i=1}^{M}(R_i-\bar{R})^2},\\
    \Delta_R=\max_iR_i-\min_iR_i,\qquad
    A_i=\frac{R_i-\bar{R}}{\max(\sigma_R,0.02)}.
\end{gathered}
\label{eq:supp_group_statistics}
\end{equation}
A group is resampled without an optimizer update if $\sigma_R<0.01$ or
$\Delta_R<0.02$. The standard deviation uses the population denominator $M$;
the advantage denominator has a floor of 0.02. The sequence-level $A_i$ is
shared by all $T_i=|y_i|$ generated tokens in rollout $i$. Here, $T_i$ is an
output length, whereas $T$ denotes the transcript.

The GRPO objective uses the following policy ratio and sampled-action
KL estimator. All token probabilities condition on the direction-specific input
and preceding generated tokens; these conditions are suppressed for brevity:
\begin{equation}
\begin{gathered}
    r_{i,t}=\exp\!\left[
    \log\pi_{\theta}(y_{i,t})-
    \log\pi_{\theta}^{\mathrm{old}}(y_{i,t})\right],\\
    \delta_{i,t}=\log\pi^{\mathrm{ref}}(y_{i,t})-
    \log\pi_{\theta}(y_{i,t}),\qquad
    \widehat{D}_{i,t}=\exp(\delta_{i,t})-\delta_{i,t}-1.
\end{gathered}
\label{eq:supp_ratio_kl}
\end{equation}
The parameter $\theta$ is $\theta_G$ or $\theta_U$ for the active direction;
$\pi_{\theta}^{\mathrm{old}}$ is the policy before the update and
$\pi^{\mathrm{ref}}$ its frozen Joint-SFT reference. We use
$\epsilon=0.2$ and $\beta=0.04$ for both directions.

GEN and UND use separate AdamW optimizers. Gradients are averaged across the
four workers and clipped to norm 1 before updating the active adapter.
CycleGRPO uses learning rate $5\times10^{-7}$, shuffled sampling, one update
per accepted group, at most 32 resampling attempts. 

\subsection{Reward Implementation and Profile-to-Instruction Bridge}
\label{sec:supp_reward_definitions}
The profile and content reward formulas are given in Section~3.4 of the main
paper. For a categorical field $k$, $s_k$ is the exact-match indicator between
the predicted and reference values. Descriptive fields use a frozen BGE-M3
encoder, with list-valued descriptions matched before aggregation.
All ten paralinguistic fields have equal weight.

English and Chinese content rewards use Whisper-large-v3 and FireRedASR2S,
respectively. WER and CER enter the reward as fractions.
For raw UTMOS22-Strong prediction $q$, quality is normalized as
\begin{equation}
    R_{\mathrm{mos}}=\operatorname{clip}\!\left(\frac{q-1}{4},0,1\right).
    \label{eq:supp_mos_reward}
\end{equation}

The deterministic bridge $\mathcal{B}$ renders the predicted profile as a
natural-language instruction and combines it with transcript $T$. Its template
is selected by the instruction category; this category dependence is suppressed
in the main-paper notation $\widetilde{X}_i=\mathcal{B}(\widehat{P}_i,T)$.
GEN then uses $\widetilde{X}_i$ and the original speaker reference $R$ to
generate tokens $\widetilde{A}_i$, which are decoded into the reconstructed
waveform $\widetilde{x}_i$.
The backward-loop $R_{\mathrm{sty}}$ compares $\widetilde{x}_i$ with $x^{*}$
using the frozen ParaMETA representation.

\section{Evaluation Protocols}
\label{sec:supp_evaluation_protocols}

\subsection{Style Accuracy Evaluators}
\label{sec:supp_style_evaluators}
For Style Acc., VoxProfile predicts gender and age, while VoxLect and
FireRedLID predict English and Chinese accents, respectively.
Emotion is assessed by a speech emotion recognition (SER) model built on
the Qwen3-Omni Captioner audio encoder and trained on large-scale emotion data.
Speaking rate is estimated using Voice-Taxonomy TEMP, and volume is
measured by mean-frame root-mean-square (RMS) amplitude.

\subsection{Instruction Match Evaluation Prompt}
\label{sec:supp_instruction_match}
Instruction Match uses Gemini-2.5-pro with the InstructTTSEval-based evaluation
prompt. The prompt is reproduced below to preserve
the evaluation criteria and requested output format.

\begingroup
\centering
\small
\begin{tcolorbox}[
  enhanced,
  breakable,
  colback=gray!5,
  colframe=green!50!black,
  title=Instruction Match Evaluation Prompt
]

You are an expert with rich knowledge of acoustics. Please describe an audio clip based on the following dimensions and judge whether the audio matches the provided description. Output \textbf{true} (matches) or \textbf{false} (does not match) on the consistency dimension, ignoring non-style factors (audio quality, naturalness, etc.).

\medskip\noindent\textbf{Evaluation Dimensions}\par\smallskip

\begin{itemize}
    \item \textbf{Gender:} Speech characteristics associated with different gender identities, including vocal fold differences and socialized speech patterns.
    \item \textbf{Pitch:} The perceived frequency of the sound, determining whether the voice sounds high or low. Typically, male voices have lower pitch and female voices have higher pitch. Relative pitch levels can be expressed based on gender, such as ``female high pitch'' or ``male deep and steady pitch''.
    \item \textbf{Speaking Rate:} The speed of speech, which often varies during a conversation. If the speaker demonstrates a specific rhythmic pattern, please specify it.
    \item \textbf{Volume:} The loudness or softness of speech, which may vary significantly. Examples include whispering, normal conversational volume, or shouting.
    \item \textbf{Age:} Infer the age group or life stage of the speaker (e.g., child, teenager, young adult, middle-aged, elderly) based on speech characteristics. If a specific stage is difficult to determine, simply indicate an approximate stage.
    \item \textbf{Accent:} A distinctive way of pronunciation reflecting geographical origin, socioeconomic background, or non-native speaker status. If the accent is sufficiently distinct, specify the dialect region as precisely as possible; otherwise, indicate a general region, such as American English, British English, or Mandarin.
    \item \textbf{Timbre \& Texture:} The tonal quality and texture of the voice, including descriptions such as sweet, raspy, deep, bright, warm, nasal, soft, harsh, or delicate. These attributes reflect physiological characteristics (e.g., vocal cord structure) and stylistic nuances, which can be used to distinguish speakers or analyze emotional/expressive tendencies.
    \item \textbf{Emotion:} The emotion expressed while speaking, which may change during speech. For example, a person might start speaking calmly but gradually become frustrated, or switch from sadness to laughter within the same sentence.
    \item \textbf{Intonation / Tone:} The emotional or attitudinal nuance conveyed through vocal inflections, including pitch variation patterns, expressing subtle nuances such as sarcasm, formality, enthusiasm, or indifference.
    \item \textbf{Personality:} Infer the speaker's overall personality based on the above speech characteristics, such as extroverted/introverted, confident, decisive, or anxious. Describe only prominent and consistent personality traits present in the speech.
    \item \textbf{Role and Scenario Performance:} Whether the role state, relationship identity, and scenario atmosphere presented by the speech match the description, such as a caregiver's restraint and exhaustion, a sense of boundaries during an argument, or professionalism in a formal broadcast scenario. Evaluate only the performance style and state that can be perceived through sound, and do not judge information that cannot be verified by audio alone (e.g., specific location, physical appearance, actions, interpersonal relationships).
\end{itemize}

\medskip\noindent\textbf{Judgment Criteria}\par\smallskip

\begin{center}
\renewcommand{\arraystretch}{1.3} % 调整行高，让表格呼吸感更好
\begin{tabularx}{\linewidth}{|l|X|}
\hline
\textbf{Judgment} & \textbf{Definition} \\
\hline
\textbf{true (Matches)} & The audio sample satisfies the main stylistic features of the description:
\begin{itemize}
    \item Primary style dimensions (e.g., gender, pitch, speaking rate, emotion) are consistent with the description
    \item No obvious deviations or conflicts
\end{itemize} \\
\hline
\textbf{false (Does not match)} & The audio sample fails to satisfy the main stylistic features of the description:
\begin{itemize}
    \item At least one key stylistic feature obviously conflicts with the description
    \item The overall auditory perception deviates from the described style
\end{itemize} \\
\hline
\end{tabularx}
\end{center}

\medskip\noindent\textbf{Precautions}\par\smallskip

\begin{itemize}
    \item When there is an obvious and objective discrepancy in a certain dimension, such as speaker gender or age conflicting with the description, judge directly as false.
    \item There is a high probability that the description obviously conflicts with, differs in degree from, or completely mismatches the audio. Do not easily trust the provided description; hold your own independent understanding of the audio first.
    \item Pay special attention to speaker gender, as it is particularly prone to being opposite to the description.
    \item Pay extra attention when words indicating degree (such as ``excited'' or ``intense'') are mentioned in the description. The specified dimension in the audio may not be as intense as described (e.g., emotions are not excited enough, pitch is not high enough, or volume is not loud enough); in such cases, judge as false.
    \item Evaluate only \textbf{stylistic consistency}, ignoring non-style factors such as pronunciation accuracy or naturalness.
    \item Use the description as the sole basis without personal subjective bias.
    \item Features not mentioned in the description have no restrictions and should not affect the judgment.
    \item When the description focuses only on a specific dimension (e.g., emotion), judgment should focus primarily on that dimension.
\end{itemize}

\medskip\noindent\textbf{Output Format Requirements:}\par\smallskip
Return a JSON object containing all fields below, replacing the placeholder
strings with audio-based descriptions. Set \texttt{Consistency} to the Boolean
\texttt{true} or \texttt{false} according to the judgment criteria; the value
shown below is illustrative.

\begin{lstlisting}[basicstyle=\small\ttfamily]
{
    "Gender": ...,
    "Pitch": ...,
    "Speaking Rate": ...,
    ...
    "Consistency": true
}
\end{lstlisting}

\medskip\noindent\textbf{Description and Audio to be Evaluated:}\par\smallskip
$<$Insert description of audio style to be evaluated here$>$
\end{tcolorbox}
\endgroup

\clearpage
\subsection{Subjective Evaluation Protocol}
\label{sec:supp_subjective_protocol}
\begin{figure}[H]
    \centering
    \includegraphics[width=0.99\linewidth,height=0.8\textheight,keepaspectratio]{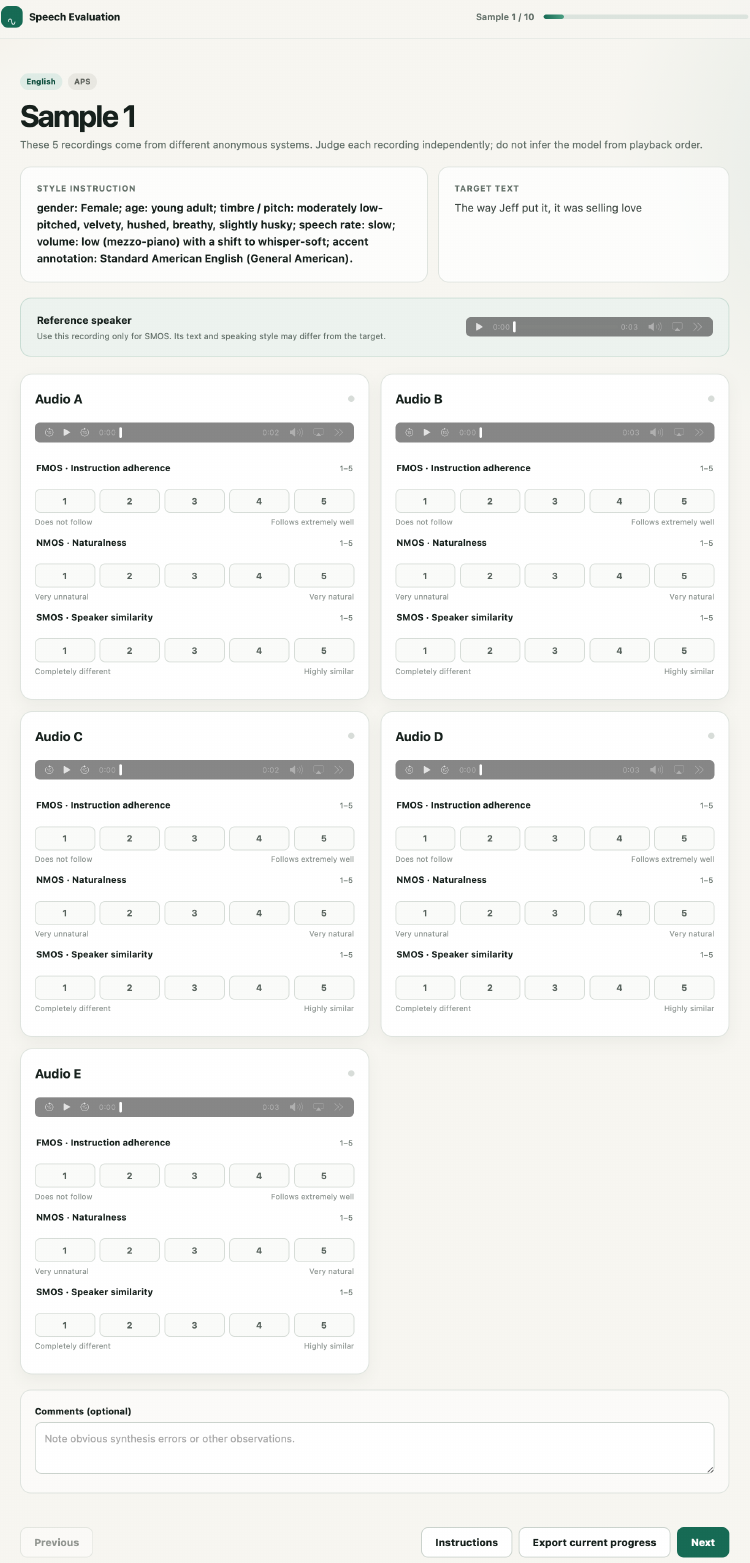}
    \caption{Listening-test interface for rating instruction/style faithfulness, naturalness, and speaker similarity.}
    \label{fig:listening_test_interface}
\end{figure}

Five listeners participated in the subjective evaluation, all with English
proficiency equivalent to IELTS 6.5 or higher. For each system, we evaluated
50 utterances, evenly split between Chinese and English (25 each), and each
utterance received ratings from at least two listeners. Samples were anonymized
and presented in randomized order. Faithfulness (FMOS), naturalness (NMOS),
and speaker similarity (SMOS) were each rated on a five-point scale.
Figure~\ref{fig:listening_test_interface} shows the evaluation interface.

\clearpage
\section{Additional Ablations}

\subsection{Stage-I SFT Ablation}
\label{sec:stage1_warmup_ablation}

\paragraph{Experimental setup.}
We compare Independent SFT (A1), which trains GEN and UND separately,
with three Joint-SFT schedules: \texttt{GEN0/UND0},
\texttt{GEN2k/UND0}, and \texttt{GEN2k/UND1k}.
The notation \texttt{GENx/UNDy} indicates the branch-specific warm-up
steps used before joint training.
Across the four configurations, we evaluate checkpoints
between $1\mathrm{k}$ and $10\mathrm{k}$ steps on a fixed set,
balanced between Chinese and English.
For each case, both GEN and UND produce six candidates under the same
reference annotations and scoring protocol.

\begin{figure}[!htbp]
    \centering
    \includegraphics[
        width=\linewidth,
        height=0.56\textheight,
        keepaspectratio
    ]{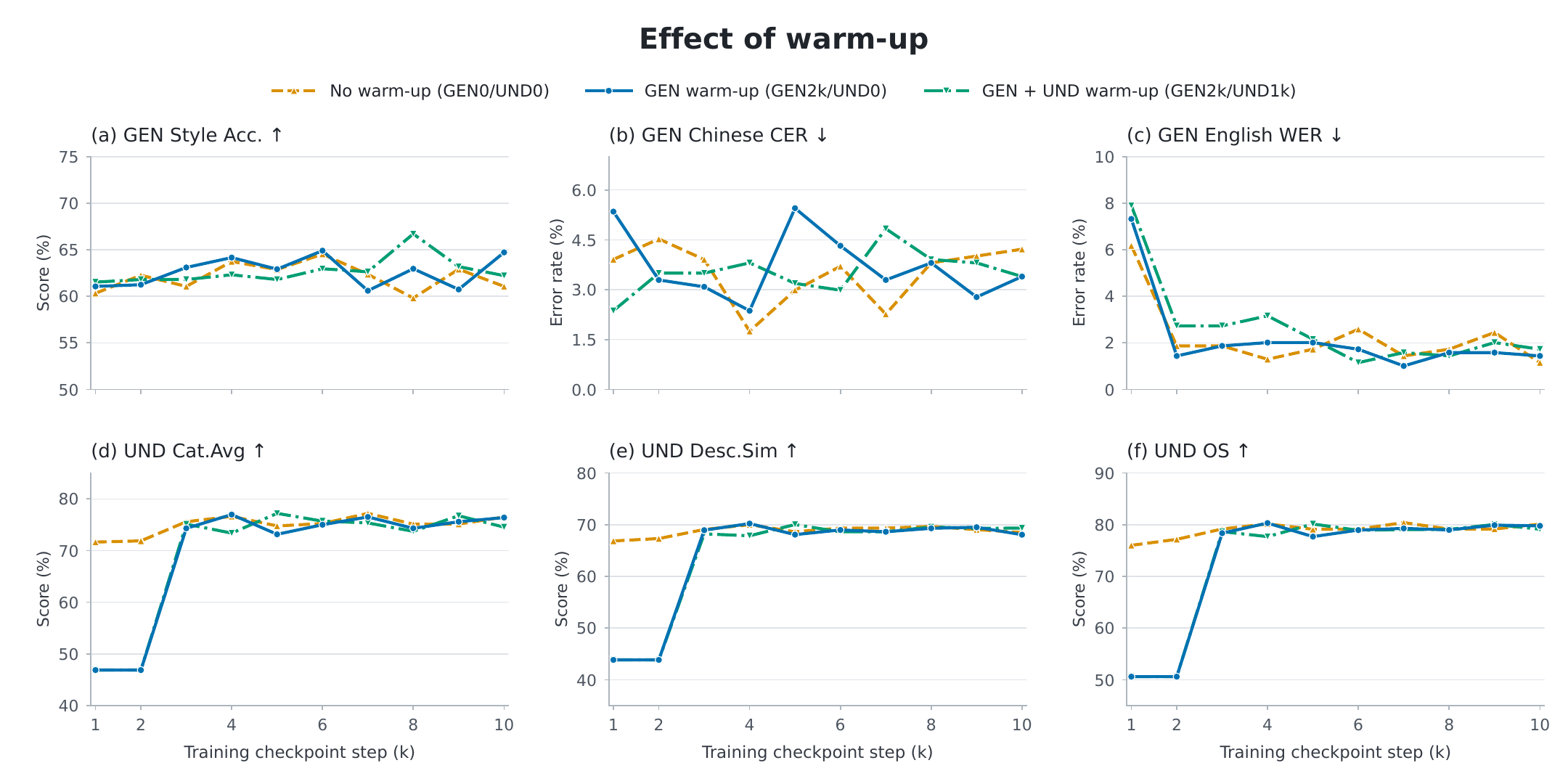}
    \caption{Joint-SFT trajectories through $10\mathrm{k}$ training steps.}
    \label{fig:sft_warmup_comparison}
\end{figure}

\begin{figure}[!htbp]
    \centering
    \includegraphics[
        width=\linewidth,
        height=0.68\textheight,
        keepaspectratio
    ]{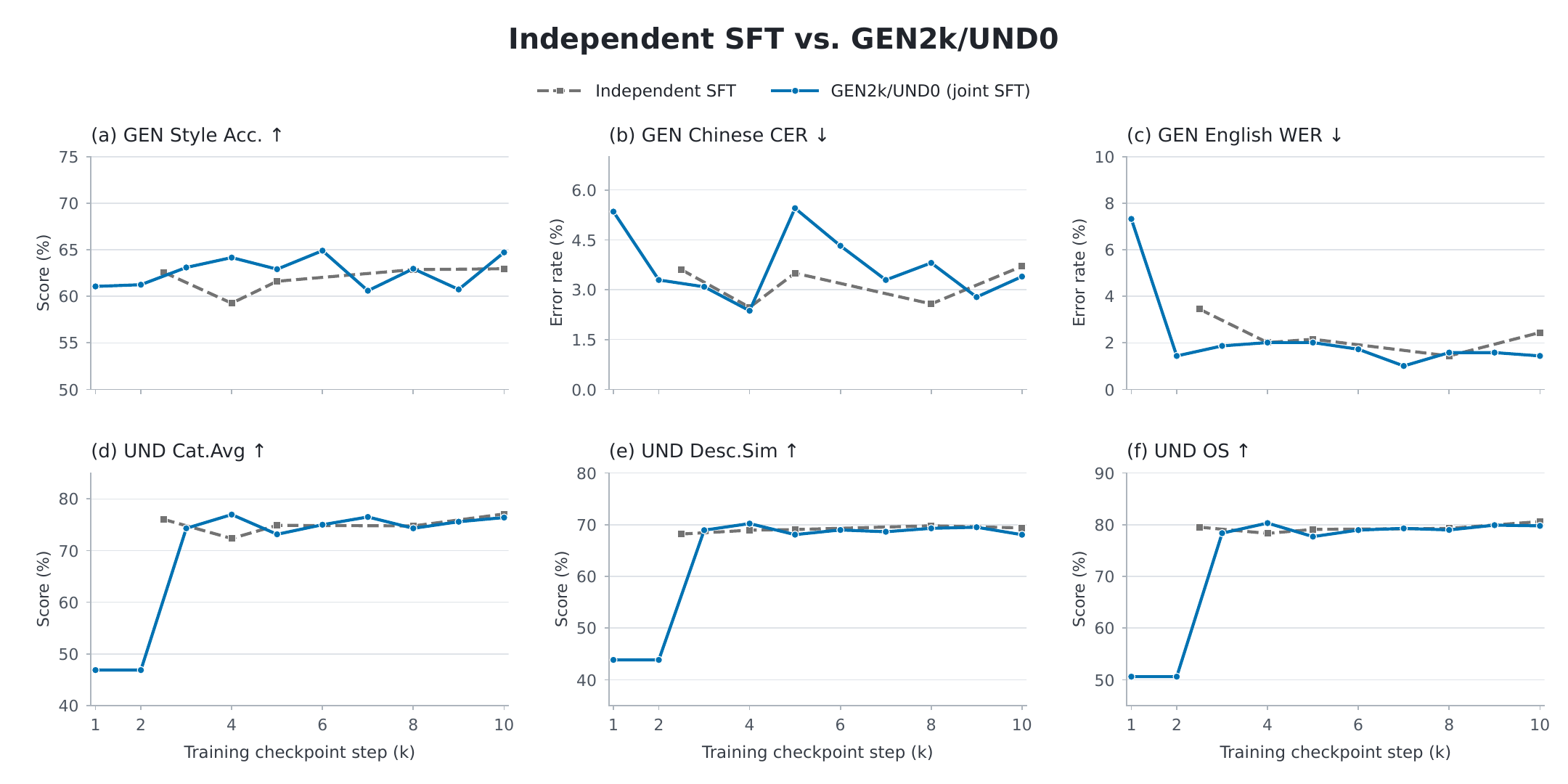}
    \caption{Comparison of Independent SFT (A1) and
    \texttt{GEN2k/UND0} through $10\mathrm{k}$ training steps.}
    \label{fig:independent_sft_comparison}
\end{figure}

\paragraph{Results and GRPO initialization.}
As shown in Figures~\ref{fig:sft_warmup_comparison}
and~\ref{fig:independent_sft_comparison},
\texttt{GEN2k/UND0} achieves the highest GEN Style Acc.\ among the four
configurations at $10\mathrm{k}$ steps ($64.72\%$).
Compared with Independent SFT, it also reduces Chinese CER from
$3.70\%$ to $3.40\%$ and English WER from $2.44\%$ to $1.44\%$,
while its UND OS is slightly lower ($79.78\%$ versus $80.61\%$).

Within \texttt{GEN2k/UND0}, the $10\mathrm{k}$ checkpoint provides higher
Style Acc.\ and lower English WER than the $4\mathrm{k}$ checkpoint,
whereas the latter performs better on Chinese CER and UND metrics.
We therefore initialize CycleGRPO from \texttt{GEN2k/UND0} at
$10\mathrm{k}$, prioritizing stronger generation control while retaining
competitive profile-recovery performance.
\end{document}